# Atomic-accuracy prediction of protein loop structures through an RNA-inspired ansatz

Rhiju Das

Departments of Biochemistry and Physics, Stanford University, Stanford, CA 94305, USA

Phone: (650) 723-5976.
Fax: (650) 723-6783.
E-mail: rhiju@stanford.edu.

# Abstract

Consistently predicting biopolymer structure at atomic resolution from sequence alone remains a difficult problem, even for small sub-segments of large proteins. Such loop prediction challenges, which arise frequently in comparative modeling and protein design, can become intractable as loop lengths exceed 10 residues and if surrounding side-chain conformations are erased. Current approaches, such as the protein local optimization protocol or kinematic inversion closure (KIC) Monte Carlo, involve stages that coarse-grain proteins, simplifying modeling but precluding a systematic search of all-atom configurations. This article introduces an alternative modeling strategy based on a 'stepwise ansatz', recently developed for RNA modeling, which posits that any realistic all-atom molecular conformation can be built up by residue-by-residue stepwise enumeration. When harnessed to a dynamic-programming-like recursion in the Rosetta framework, the resulting stepwise assembly (SWA) protocol enables enumerative sampling of a 12 residue loop at a significant but achievable cost of thousands of CPU-hours. In a previously established benchmark, SWA recovers crystallographic conformations with sub-Angstrom accuracy for 19 of 20 loops, compared to 14 of 20 by KIC modeling with a comparable expenditure of computational power. Furthermore, SWA gives high accuracy results on an additional set of 15 loops highlighted in the biological literature for their irregularity or unusual length. Successes include *cis*-Pro touch turns, loops that pass through tunnels of other side-chains, and loops of lengths up to 24 residues. Remaining problem cases are traced to inaccuracies in the Rosetta all-atom energy function. In five additional blind tests, SWA achieves sub-Angstrom accuracy models, including the first such success in a protein/RNA binding interface, the YbxF/kink-turn interaction in the fourth 'RNA-puzzle' competition. These results establish all-atom enumeration as an unusually systematic approach to *ab initio* protein structure modeling that can leverage high performance computing and physically realistic energy functions to more consistently achieve atomic resolution.

**Introduction**

Atomic-resolution prediction of protein three-dimensional structure is a biophysical problem with fundamental implications for the structure determination and rational engineering of complex biological systems [1-4]. Recent years have seen major successes in modeling protein structure through the *in silico* optimization of all-atom energy functions [2-4]. However, as assessed in blind trials, these computational algorithms achieve atomic accuracy only in favorable cases [5-7] or when guided by experimental data [8-10], even with the application of new kinds of specialized supercomputers [11]. Even relatively short sequences, such as loops involved in catalysis or in binding of drugs or macromolecule partners, present a massive number of possible conformations that cannot be exhaustively searched [1,12]. Most available methods thus make use of coarse-grained search phases using knowledge-based potentials or approximate filters to reduce the number of energy minima that need to be searched [2,4-6,8,10,12].

Recently, a conceptually distinct approach to modeling macromolecule structure has arisen from efforts to predict complex RNA structures in all-atom detail [13-16]. A working hypothesis, called a 'stepwise ansatz', posits that native biopolymer structures can be built through the systematic step-by-step addition of one residue at a time. When integrated via a dynamic-programming recursion, this ansatz permits the enumeration of a physically realistic subspace of molecular conformations at all-atom resolution, and was implemented as a stepwise assembly (SWA) protocol in the Rosetta framework. In a comprehensive benchmark, this *ab initio* method consistently eliminated conformational sampling bottlenecks and solved RNA loops and motifs at high resolution. The method has furthermore been successful in blind tests [13,14], but in all cases has required the expenditure of significant computational power (thousands of CPU-hours). When coupled to limited experimental data, SWA can be accelerated and is enabling the determination of difficult NMR structures from limited RNA chemical shift data ([17,18]; Sripakdeevong, P. and RD, submitted) and the automated correction of errors in fitting RNA coordinates into crystallographic density maps (the ERRASER method [15,18,19]).

Given its advantages over prior RNA modeling approaches and its assurance of complete sampling, stepwise assembly also holds promise for difficult problems in protein structure prediction. This study presents the first application of SWA to proteins, focusing on loop modeling. Protein loop modeling problems arise frequently in comparative modeling, designing new proteins, and solving or refining protein folds with limited crystallographic or NMR data, including weakly populated ('invisible') states [20-24]. When knowledge of the side-chains inside and outside the loop is erased, this problem has remained generally difficult, as it involves searching dozens of backbone torsions and hundreds of side-chain torsions to achieve a precise 'lock-and-key' fit between the loop and the surrounding protein [21,22]. Despite important methodological advances in recent years, inefficient conformational sampling has continued to be a general bottleneck in protein loop modeling [21,22,24].

To address this conformational sampling bottleneck, this article describes the import of stepwise assembly from RNA modeling to protein loop structure prediction in the Rosetta framework (Figure 1). The resulting SWA method is tested in a benchmark of thirty-five protein loops, including numerous cases that have been challenges for prior approaches. The benchmark includes loops with conformations ranging from corkscrews to hairpins, complex motifs such as *cis*-Pro touch turns, loops that thread through tunnels formed by the surrounding protein, and segments of unprecedented length. These cases are solved *ab initio* by SWA with sub-Angstrom accuracy, albeit at the expense of thousands of CPU-hours per loop. Additional atomic-accuracy results from five blind predictions, including a protein/RNA complex, give further support to the stepwise ansatz and its Rosetta SWA implementation. Analogous to recent successes of RNA modeling in structural biology [13-16,19], all-atom conformational enumeration may be useful for systematically dissecting protein structure and dynamics in practical scenarios with limited experimental data.

**Results**

**Difficulty of loop structure prediction with atomic accuracy**

Protein loops provide well-defined problems for structure prediction, as exemplified in Figure 2 by a 12-residue segment of a flavoenzyme (sequence DPHSNTRTDEYG; residues 203-214 in PDB entry *1oyc*), an unsolved case in recent studies [21,22]. Compared to regions of regular secondary structure, protein loops contain similar numbers of hydrogen bonds (1.1 per residue, on average; see $N_{HB}$ in Table 1), but contain few 'regular' α-helix-like or β-sheet-like backbone-backbone hydrogen bonds (only two in the *1oyc* case). Furthermore, loops make few interactions with the non-polar core, arguably the best modeled region of protein structures [2,25,26], and present few non-polar side-chains. For example, nearly all of the *1oyc* loop residues (Table 1) have charged or polar side-chains, which require atomically precise positioning to form hydrogen bonds; and the other residues are glycine and proline, which each have unique conformational properties [27].

These factors have made the *1oyc* test case difficult for state-of-the-art methods, including the Protein Loop Optimization Protocol (PLOP), which uses backbone-only hierarchical loop build-up followed by molecular mechanics force field refinement [20,21]; fragment-assembly modeling with all-atom refinement in the Rosetta framework [22,24]; and a more powerful Rosetta algorithm based on Monte Carlo sampling with exact kinematic loop closure (KIC) [22]. To further test the limits of state-of-the-art methods, Rosetta KIC modeling has been repeated herein with several enhancements, including generation of 16,800 models (greater than 16-fold times the computational expenditure of prior work [22]), *cis* proline sampling, and more stringent chain closure. Even in these more extensive calculations (Table 1), the best of five lowest energy cluster centers for the *1oyc* loop achieved a C$\alpha$ RMSD accuracy of 3.03 Å, significantly worse than RMSD values less than 1.0 Å associated with atomic accuracy solutions [20-22]. Indeed, none of the KIC models attained better than 1.68 Å RMSD to the crystallographic loop. Nevertheless, as shown below, near-native loop configurations with substantially lower all-atom Rosetta energies exist for the loop. These results illustrate the difficulty of high-resolution protein loop structure prediction, even with the application of large amounts of computational power.

**Importing a stepwise ansatz from RNA modeling**

Most current loop modeling approaches that seek atomic resolution share a seemingly necessary working approximation: an initial search phase using a reduced representation with simplified or no side-chain atoms. Such coarse search phases avoid the complexity and large number of local minima inherent to all-atom representations, but fail to capture hydrogen bonds and non-polar packing interactions involving side-chain atoms, which are pervasive (see $N_{SC}$ in Table 1). A strategy that ensures sufficient sampling, or even enumeration, of such interactions in all-atom detail would appear desirable. However, even if constrained to fixed bond lengths and angles, planar *trans*-peptide bonds ($\omega=180°$), crystallographically observed ($\phi,\psi$) combinations, and closed loop geometries, sampling each amino acid at sub-Angstrom resolution requires at least tens of backbone conformers (see SI Methods), resulting in at least $10^{12}$ backbone torsional combinations for a 12-residue loop. Subsequent side-chain optimization on these backbones would require tens of millions of CPU-hours [28]. Furthermore, the combinatorial space becomes exponentially larger with increasing loop length.

Recent work in three-dimensional RNA modeling [13] has suggested a distinct strategy for conformational enumeration, based on the following working hypothesis, or ansatz: the native conformation of a loop can be built up progressively in small steps such that each intermediate partial conformation is itself a well-packed, low-energy configuration (Figure 1). As implemented herein, the steps involve adding a single residue to either a fragment growing from the N-terminus or the C-terminus of the loop (Figs. 1a & 1b) and bridging the two fragments when they arrive within three residues of each other (Fig. 1c).

Each step involves three sub-steps: enumeration of backbone torsions for the new residue and any sequence-adjacent loop residues; optimization of the new and nearby side-chains; and minimization of the entire loop and surrounding protein side-chains (see Methods for detailed descriptions of residue capping, torsions sampled, chain closure algorithm, and example Rosetta command-lines). An ensemble of 400 models is retained and typically includes models within 5 $k_BT$ of the lowest energy model at each stage, mimicking a thermal ensemble (here, a Rosetta unit is taken to be approximately one $k_BT$ [29]). Each step is enumerative and deterministic except the local side-chain optimization. While this side-chain optimization problem can be solved exactly [30,31], SWA relies on Rosetta's stochastic one-at-a-time sampling [28] for speed; the resulting side-chain search appears near-optimal, as independent trials of the entire stepwise calculation gave lowest-energy final models with similar conformations (<0.5 Å RMSD) and energies (within 1-2 $k_BT$). This protocol has been coded into the Rosetta framework as a stepwise assembly (SWA) algorithm.

As an example, Figs. 2a-k illustrate the step-by-step building of a near-native *1oyc* loop. For the first five residues (N-terminal 203-206 & C-terminal 214; Figs. 2a-e), conformations within 0.6 Å of the crystallographic loop are observed in the ensembles of 400 lowest energy models for the growing chain, although not as the very lowest energy model (see SI Fig. S1 for full energy vs. RMSD plots). The features that stabilize these near-native conformations all involve side-chains: a side-chain hydrogen bond from D203, a backbone-to-side-chain hydrogen bond from G214, non-polar packing by the P204 prolyl ring, separate packing interactions by H205, and again a side-chain hydrogen bond from S206. In prior approaches, initial search phases that omit or coarse-grain side-chains for computational simplicity would necessarily miss these key, atomic-level details. The SWA modeling of the *1oyc* loop then continues through additional rebuilding steps from the N- and C- termini and chain closure across residues D211-E21 (Figs 2f-k). The final lowest energy model for the full loop achieves a C$\alpha$ RMSD of 0.39 Å, with all backbone and side-chain hydrogen bonds recovered with atomic accuracy (Fig. 2l). Reaching this final lowest energy solution requires intermediate structures that are not the lowest energy at their steps (SI Fig. S1), underscoring the need for keeping a full ensemble of models during the build-up procedure.

The preceding description outlined one potential residue-by-residue build-up path, but it is not known *a priori* which path, if any, will lead to the native loop. It is therefore critical to search through all possible build-up paths. This task is simplified by the observation that each path shares most of its sub-paths with other paths. The full build-up can thus be accomplished through a dynamic-programming-style recursion similar to methods in modeling RNA secondary structure [32] and, more recently, tertiary structure [13]. Briefly, each intermediate in the SWA build-up can be indexed by the ends of the N-terminal and C-terminal fragment of the loop; the ensembles to be modeled can thus be laid out schematically in a two-dimensional matrix (Fig. 1d). Members of each ensemble are computed by applying the build step to all relevant models in a previous ensemble (arrows to immediately neighboring matrix elements in Fig. 1d), followed by clustering of the resulting ensemble and retaining the lowest 400 models. The resulting number of build steps grows quadratically, rather than exponentially, in the number of residues in the loop and can be carried out with approximately one day of computation on a 200-CPU cluster (5,000 CPU-hours). [A simpler and faster recursion (Fig. 1e) that assumes limited interaction between C- and N- terminal fragments of the loop has also been implemented and tested (see Methods and Table 2), and can be carried out in advance of the full recursion.] The importance of search path in sampling diverse loop conformations is illustrated by the large differences in models with comparably low energy achieved by following different paths, as shown in Fig. 3. Nevertheless, for the *1oyc* case, the high-accuracy 0.39 Å conformation (Fig. 2l) remains the lowest energy loop after following all rebuild paths (Figs. 2m and 3b). This SWA model's energy is substantially lower than any conformation produced by Rosetta KIC Monte Carlo modeling (Fig. 2m).

The SWA method posits that the experimentally observed conformation of a protein loop is part of a low energy subspace that can be enumerated through the stepwise, locally optimal building of small subsegments. This working hypothesis – the stepwise ansatz [13] – appears feasible for native macromolecule conformations, in which nearly every residue makes precise, atomic-level interactions with other residues (see, e.g., Figs. 2a-k, and $N_{contact}$, $N_{out}$, $N_{SC}$, and $N_{HB}$ entries in Table 1). As with prior RNA methods [13], however, general confirmation of the ansatz requires extensive empirical tests on a wide range of protein loop structures, described next.

**Benchmarking the SWA algorithm**

To test the stepwise ansatz, its Rosetta stepwise assembly (SWA) implementation was used to carry out structure prediction on a benchmark set of thirty-five protein loops.

Twenty of these cases were 12-residue loops used previously to test PLOP and Rosetta approaches [21,22]. Fifteen additional cases with lengths between 8 and 24 residues were chosen from studies of loop modeling and classification [23,27,33-35] that highlighted their complex but well-defined geometries (see Table 1 for full descriptions).

First, examining the subset of twenty loops used in prior PLOP and Rosetta studies permitted direct comparison of SWA to these state-of-the-art methods (see also SI Table S1 and SI Fig. S2). Here, the median C$\alpha$ RMSD accuracy (lowest energy structure) was 0.64 Å, consistent with sub-Angstrom accuracy and lower than values for prior methods: 2.3 Å (PLOP [21]), 1.2 Å (PLOP with surrounding side-chain optimization [21]), 2.1 Å (Rosetta fragment assembly [22]), 1.0 Å (Rosetta KIC [22]), and 0.84 Å (Rosetta KIC repeated herein with computational power comparable to SWA calculations; Table 1 and SI Table S2). The SWA method thus outperforms prior loop modeling approaches.

In 19 of the 20 cases, at least one of the five lowest energy SWA models achieved sub-Angstrom accuracy. For comparison, KIC modeling achieved this level of accuracy in fewer cases, 14 of 20 (Table 1).[*] The high accuracy SWA predictions that were intractable to previous PLOP and/or Rosetta approaches included the *1oyc* case described above (Fig. 2l) as well as a loop containing both a *cis* proline and *trans* proline from ZipA (*1f46*, Fig. 4a), a loop with a 'corkscrew' fold in tetanus toxin C (*1a8d*; Fig. 4b), a highly extended loop from an immunoglobulin domain involved in neural cell adhesion (*1cs6*, Fig. 4c), and a hairpin-like loop from a bacterial esterase (*1qlw*; Fig. 4d). The residue-by-residue paths that achieved the high accuracy SWA models were different for each case (pathway traces in Fig. 4), underscoring the necessity of following all build-up paths.

---

[*] A smaller number of cases (11 of 20) was reported as 'solved' by KIC in prior work, which used less computational power (1000 CPU hours), did not sample *cis* prolines, and reported RMSD for the very lowest energy model.

For the 15 additional test cases, SWA gave somewhat lower performance, as expected, given that these cases were selected for their complexity. SWA achieved median Cα RMSD accuracies of 1.4 Å (lowest energy structure) and 0.89 Å (best of five lowest energy structures). SWA modeling again outperformed KIC modeling overall, although not by as much as in the first 20-loop benchmark [1.9 Å (lowest energy) and 0.94 Å (best of five)]. In 8 of these 15 complex loop tests, SWA returned at least one of five lowest energy models with sub-Angstrom accuracy (Table 2; see also SI Fig. S3). High-resolution structures were recovered for segments that contained *cis*-Pro touch turns [35] (*1msp*; Fig. 4e), that threaded through 'tunnels' formed by other side chains in α-lactalbumin [21] (*1alc*, a highlighted problem case for PLOP [20,21]; Fig. 4f), and that bound inhibitors in papain [36] (*1ppn*; Fig. 4g). KIC modeling also achieved sub-Angstrom accuracy in 8 of 15 test cases, but not in all the same cases (see below).

The most striking SWA models involved loops with long lengths (Figs. 4h & i). Formally, the exponential growth of possible conformations with loop size makes a 24-residue loop puzzle substantially more difficult than a 12-residue loop (with approximately $10^{12}$–fold more accessible conformations). However, the stepwise ansatz underlying the SWA method constrains sampling to a subspace that requires only 4-fold more steps to search. For three of the six cases with lengths greater than or equal to 18 residues, the SWA method achieved sub-Angstrom accuracy. Modeling with such accuracy included two 24-residue cases. One involved a mixture of irregular, helix, and strand segments in a bacteriophage head protein (*1c5e*; Fig. 4h), and another involved a long loop threading through the center of a lipase domain (*1thg*; Fig. 4i). For these loops, extensive KIC modeling runs (Table S2) failed to achieve any models at any energy with RMSD accuracy better than 2.0 Å (SI Fig. S4). These results illustrate the effectiveness of the stepwise ansatz in reducing the vast conformational space of a protein segment into a physically realistic subspace that can be systematically searched with available computational resources.

**Problem cases and the Rosetta energy function**

Overall, 27 of the 35 loop puzzles were solved with atomic accuracy by the SWA method, taking into account the five lowest energy models. Most of the residual problems appeared due to inaccuracies in the assumed energy function, analogous to observations made for SWA modeling of RNA loops [13]. For example, even amongst the 27 success cases, the best of five lowest energy conformations – but not the very lowest energy conformation – achieved sub-Angstrom accuracy, suggesting imperfect energy function discrimination amongst these low energy states. Further evidence for energy function problems came from SWA problem cases. For six of the eight cases in which sub-Angstrom accuracy was not achieved, the SWA approach uncovered non-native models with energies within 3 $k_BT$ of optimized crystallographic loops (SI Table S2; four cases were within 1 $k_BT$). Interestingly, in two of these cases (*1arp* and *1huw*), KIC modeling outperformed SWA modeling in terms of RMSD but gave significantly worse Rosetta energies (> 10 $k_B$T; SI Table S2 and Figure S5). This comparison suggests that SWA's strong optimization of the Rosetta all-atom energy function – apparently quite inaccurate in these two cases – prevented it from sampling the full diversity of conformations discovered by the KIC method in its low-resolution stage. These two loops – as well as two of the eight problem cases in which SWA gave significantly worse energies than the experimental loop – were solvent exposed and making few non-polar interactions. Future improvements in the Rosetta energy function, particularly in its highly oversimplified solvent model [37,38], may better guide SWA modeling in early stages to partial loop conformations that give better all-atom energies and/or accuracies.

**Blind tests**

Blind trials with sequences of previously unknown structure provide important tests of structure prediction methods. For this study, SWA was tested on five such cases, from two sets of problems. Four problems involved the loops from a 275-residue crystallographic model of an all-α protein complexed to a long helix, recently solved by Weis and collaborators (Stanford University) and not released outside their research group.[†] The closest previously solved structure exhibited low sequence identity to the target (26% over a 210-residue alignable region); and analogs of the loop regions either did not exist (loop A) or were different in sequence at all 12 positions (loop B) or at 10 positions (loops C and D) in the homologous structure. The Weis group provided a starting structure with all of these loops and all side-chains removed, and *ab initio* SWA models were generated for the four loops. Upon un-blinding, SWA models gave sub-Angstrom recovery in all four cases (Table 2; Figs. 5a-d), including a 0.90 Å model of loop A, which formed an irregular lasso around the protein's binding partner (Fig. 5a). In three of these four cases, KIC modeling also achieved sub-Angstrom accuracy (Table 2).

---

* The Weis group has requested that the identity of the protein remain confidential during review, as the structure has not yet been published.

As a fifth blind test, SWA models were generated for a loop of protein YbxF [39] that interacted with the SAM-I riboswitch RNA, an RNA/protein target that constituted the fourth 'RNA-puzzles' trial [14]. This problem was more challenging than those above, as the starting structure was not a crystallographic model but instead a comparative model [40% sequence identity to template *2fc3*] based on threading with HHPRED and Rosetta [5,40]. SWA modeling of the loop, including nearby RNA atoms as potential interactors, gave a conformation 2.0 Å C$\alpha$ RMSD away from the loop in the starting comparative model. Nevertheless, this loop agreed with the conformation in the subsequently released structure (*3v7e*; Fig. 5e) at 0.53 Å C$\alpha$ RMSD. These results demonstrate the utility of the SWA protocol in a complex structure prediction context. An exact comparison with KIC was not possible here due to the lack of a coarse-grained RNA/protein interaction potential in Rosetta; however, KIC modeling of the protein loop without the RNA also returned a sub-Angstrom accuracy loop (0.8 Å C$\alpha$ RMSD), albeit as the second lowest energy model.

## Discussion

**A novel and systematic strategy for protein structure modeling**

This article has presented a strategy for protein structure prediction that achieves atomic accuracy on the majority of loop modeling targets through a systematic all-atom enumeration. Several of these targets were difficult or intractable with prior approaches. The main innovation herein is a stepwise ansatz imported from RNA structure modeling. This working hypothesis posits that realistic loop structures are reachable via the residue-by-residue building of partial conformations that are themselves well-stabilized by precise hydrogen bonds and non-polar packing interactions. This ansatz underlies a Rosetta stepwise assembly (SWA) protocol and is supported by tests of the SWA algorithm on forty loop puzzles, including twenty shared with prior loop modeling benchmarks, fifteen more difficult loop cases, and five blind tests. In the majority of cases (32 of 40), including loop puzzles of unprecedented length and all the blind tests, the SWA method achieved sub-Angstrom accuracy.

The stepwise assembly protocol is novel in protein modeling studies: while prior efforts have proposed the build-up of short peptides or lattice models [21,41-44], the SWA method herein provides a complete enumerative protocol without coarse-graining, stochastic search, or other approximations. This study has demonstrated that such calculations are achievable for a 12-residue protein loop in approximately 5,000 CPU-hours, readily accessible with modern parallel computing clusters. This expense is greater than Monte Carlo or refinement-based approaches, which, in favorable cases, can recover loop conformations in hundreds of CPU-hours or less [21,22], including some of the challenges considered herein. Nevertheless, many complex loops remain unsolvable by these prior approaches, even with the expenditure of massive computational power (Table 2 and Supporting Information Table S1). Thus, for the general case, the computational expense of SWA may be worthwhile. Future optimizations in continuous minimization of protein configurations and in criteria to prune build-up paths, are expected to accelerate the method, as well as incorporation of sparse experimental constraints (see below).

**Scaling with problem size**

Given that SWA carries out a systematic residue-by-residue enumeration, it is perhaps surprising that it remains computationally efficient for long loops. For a protein segment with length of $N$ residues, the formal size of the conformational space scales exponentially with $N$; the actual experimental folding times of proteins do not scale so poorly, implying the general existence of folding intermediates or pathways rather than a random walk search [1]. In SWA, the number of steps required to build a protein loop does scale efficiently (in polynomial time) with $N$. The resulting efficiency permitted the atomic resolution recovery herein of loops with lengths up to 24 residues, whereas prior work tackled loops no longer than 12 residues (see, e.g., [21-23]). The scaling efficiency of SWA suggests that *ab initio* buildup of full proteins, and not just loops, should be possible; preliminary work has demonstrated the feasibility of modeling mini-proteins with lengths of 30 residues [33].

**Implications for protein folding**

The general success of the SWA modeling method suggests that the lowest-energy *in silico* pathways of the calculation (see, e.g., Figs. 3 & 4) may represent actual *in vitro* folding pathways of a protein loop. In this case, the *in silico* 'instantiation' of each protein residue during SWA might be though of as the *in vitro* formation of a fixed structure by that residue, which was originally in a random-coil or transiently structured ensemble of conformations. Development of a kinetic model from the SWA-calculated energy landscape (analogous to efforts in RNA secondary structure [45]) as well as increases in spatial and time resolution of experimental single-molecule approaches to protein folding may be able to test these predicted folding pathways.

**Implications for practical structural biology**

The SWA algorithm for protein loop modeling presented herein may have unique and practical uses in several areas of structural biology. In *ab initio* structure modeling problems, as discussed above, SWA offers the potential for consistent sub-Angstrom accuracy as well as a powerful (if computationally demanding) tool for uncovering deficiencies in the Rosetta energy function, as in the solvent-exposed loops described above (see also [33]). In addition, there are numerous practical applications of *ab initio* modeling that do not rely on a 'perfect' energy function and instead make use of limited experimental data to break degeneracies. These problems include the high-resolution fitting of coordinates into low-resolution electron density maps [8], the solving of NMR structures from sparse chemical shift data [9,46], and the determination of 'invisible state' structures from both NMR and crystallographic approaches [47,48]. As with analogous RNA problems in which SWA is proving to be uniquely powerful [15,16,18,49], the protein SWA method herein is expected to be substantially accelerated in these use cases due to the inclusion of experimental constraints and to give models with a particularly high level of confidence due to the algorithm's guarantee of enumeration. Finally, efforts to design protein interfaces and enzyme active sites often encounter sampling bottlenecks due to the difficulty of simultaneously optimizing side-chain identity, side-chain conformation, and new backbones [3]. If expanded to include residue-by-residue side-chain identity optimization, the stepwise ansatz may offer an efficient working hypothesis for designing such functional molecules by enumeration.

**ACKNOWLEDGMENTS**

This work has benefited from sharing code and ideas with members of the Rosetta community, and discussions with P. Sripakdeevong, F. Cochran, and members of the Das group. The author thanks D. Mandell and T. Kortemme for the 20-loop benchmark files; W. Weis, J. Caldwell, S. Dobbins, and C. Peterson for preparing the starting structure for the four-loop blind puzzle; and J. Cruz, E. Westhof, and A. Ferré-d'Amaré for organizing the YbxF RNA-puzzle. Calculations were carried out on the BioX$^2$ cluster and TeraGrid/XSEDE resources.

**METHODS**

Stepwise assembly (SWA) was implemented in C++ in the Rosetta codebase and is available in Rosetta release 3.5, free to academic users at http://www.rosettacommons.org. Descriptions of the sampling method, the directed acyclic graphs for entire calculations, and explicit command-line examples, are given in the following sections.

*Stepwise Assembly (SWA)*
A diagram of the entire stepwise assembly (SWA) calculation is given as a directed acyclic graph (DAG) laid out in the style of a dynamic programming matrix in Figure 1d. Given a crystallographic loop to be built *de novo* from residue $k$ to residue $l$, each stage of stepwise assembly involved creating models of the loop with an N-terminal fragment built forward from $k–1$ to residue $i$ and a C-terminal fragment built backward from residue $l+1$ to $j$. Each stage could thus be indexed with the two residue positions ($i,j$). The SWA calculation proceeded recursively from stages with short fragments built into the structure towards models with longer fragments, i.e., $i$ increasing from $k–1$ or $j$ decreasing from $l+1$. This building corresponds to movement from the top-right to the bottom or left, respectively, in Fig. 1d. The SWA calculation involved five basic kinds of steps (see next section for example Rosetta command lines):

*(1) Pre-packing of the starting model*
The first step was a 'pre-packing' of the side-chains of the starting model with no loop atoms, corresponding to the top-right corner ($k–1$, $l+1$) of the DAG in Fig. 1d. This step was necessary as our starting models contained no side chains as well as no loop atoms. This prepacking stage thus placed initial side-chains (constructed with Rosetta ideal bond lengths and angles) using the Rosetta *packer*. This pack optimized the rotamers using simulated annealing, after precomputing pairwise energies between all potential side-chain rotamers [28]. After packing, the side-chain torsions were subjected to the non-monotone Armijo variant of Broyden-Fletcher-Goldfarb-Shanno (BFGS) minimization using the Rosetta *minimizer* [38,50]. All non-loop backbone degrees of freedom were held fixed here and in all later stages of the SWA calculation.

*(2) Addition of single residue to N-terminal fragment of the loop*
The core computation in SWA is the addition of a new residue to a model and enumeration of its backbone conformations. For additions to the *N*-terminal fragment, this step took models from stage ($i–1, j$) to ($i, j$) (downward arrows in Fig. 1d). The newly added residue included a methylamide group at the C-terminus, simulating the peptide connection to the next residue (including the next backbone amide; Fig. 1a). The enumeratively sampled degrees of freedom were backbone torsions for both the added residue ($\phi_i$ and $\psi_i$) and the previous, adjacent residue ($\phi_{i-1}, \psi_{i-1}$), permitting the discovery of configurations in which the dipeptide segment is stabilized by interactions by the new residue without requiring interactions at the previous residue. (For the initial loop residue, $i = k$ the previous residue torsions were fixed and not sampled). For each $\phi$ or $\psi$, the sampling was a grid search from $-180°$ to $180°$ in $20°$ increments. To keep only sterically realistic backbones, configurations in which a residue's ($\phi$, $\psi$) gave Rosetta *ramachandran* score greater than 0.8 Rosetta units were discarded. The $\omega$ torsion was assumed to be $180°$ (*trans* configuration), except for residues that preceded prolines, which were also sampled at $0°$ (*cis*). The number of backbone combinations varied from tens to several thousand (for segments involving glycine residues and/or residues that preceded proline).

For each combination of backbone torsion angles, the side-chains of the loop and its surroundings were optimized. Analogous to calculations in protein-protein docking [24,51], the side-chain optimization was focused on the two residues whose backbones were sampled (*i* and *i*–1) and their potential neighbors (the *neighbor_list*, determined based on C$\beta$ locations by Rosetta scoring). The side-chain sampling was carried out with the Rosetta *rotamer_trials* algorithm. (Runs using the more computationally intensive *packer* algorithm gave indistinguishable results; the problem involves few side chains and the optimum is found easily.) The searched side-chain rotamers included those listed in the backbone-dependent Rosetta rotamer library as well as additional rotamers with $\chi_1$ and $\chi_2$ shifted by $\pm 1$ standard deviation from the standard rotamer values. The discreteness of the backbone grid search and rotamer library can penalize favorable side-chain interactions due to minor clashes or slightly imperfect hydrogen bonds. Therefore, the energy function for side-chain optimization was modified from the current standard Rosetta all-atom energy function (*score12*) to include a lower weight on *fa_rep* (Lennard-Jones repulsion; 0.10 instead of 0.44), a higher weight on *hbond_sc* (side-chain/side-chain hydrogen bond strength; 3.1 instead of 1.1), and no attenuation of hydrogen bond strength at solvent-exposed residues [38,52].

After enumerative backbone sampling and side-chain optimization, models were clustered as follows. In order of energy, starting with the lowest energy model, the RMSD of each model to all lower energy clusters was computed; this RMSD value was calculated over N, C, C$\alpha$, and O atoms at the rebuilt residues *i*–1 and *i* (all other residues shared the same backbone configuration). If the RMSD value to any lower energy clusters was less than a fine cutoff (0.10 Å), the model was considered too close to an existing representative and discarded; otherwise the model seeded a new cluster. The lowest energy 400 models after clustering were carried forward to minimization.

Minimization involved backbone torsions ($\phi$, $\psi$, and $\omega$) at the sampled residue and torsions $\chi$ for all neighboring side-chains (Rosetta ideal bond lengths and angles were assumed as fixed throughout the SWA procedure). This torsional optimization was performed with the non-monotone Armijo variant of BFGS minimization in the Rosetta *minimizer* [38,50]; the energy function was the current standard Rosetta all-atom energy function *score12*. The models were clustered as described above, and saved to disk.

*(3) Addition of single residue to C-terminal fragment of the loop*
The 'prepending' of a new residue *j* to the C-terminal loop fragment (leftward arrows in Fig. 1b) was analogous to the step appending to the N-terminal fragment above. An acetyl group at the N-terminus of *j* simulated a peptide connection to the previous residue (Fig. 1b). Residues *j* and *j*+1 were subjected to enumerative backbone search, side-chain packing, model clustering, torsional minimization, and final clustering as above.

*(4) Chain closure*
For the final stages of SWA assembly, N-terminal and C-terminal fragments were bridged to form continuous loops with ideal backbone bond lengths and angles (Figure 1c), and the entire resulting loops were subjected to continuous minimization. Chain closure attempts were carried out for all models in which the number of gap residues between the N-terminal and C-terminal fragments was 1, 2, or 3 [that is, for models from stage $(i, j)$ where $1 \leq (j - i - 1) \leq 3$].

As preparation for chain closure, the N-terminal gap residue *i*+1 was appended to the N-terminal fragment, and other gap residues *i*+2 to *j*–1 were prepended to the C-terminal fragment. The $\phi$ and $\psi$ torsion angles of the first gap residue *i*+1 were sampled by grid search as above; and, to attempt chain closure, backbone torsions for 'bridge' residues *i*+2 up to *j*–1 were subjected to 1000 cycles of cyclic coordinate descent [CCD; *fast_ccd_loop_closure* in Rosetta [28,53]]. CCD was applied to all $\phi$ and $\psi$ torsion angles at bridge residues; on the $\psi$ torsion immediately preceding the first bridge residue (here on the other side of the chainbreak); and on the $\phi$ torsion immediately after the last bridge residue. Any models with chain closure RMSD (see below) less than 1.5 Å were then passed forward to side-chain optimization, clustering, and then continuous torsional minimization as above, except that all loop side-chains (not just newly built residues) and all loop backbone torsions were optimized in these steps, along with neighboring side-chains from the surrounding protein scaffold. The last full-loop torsional minimization offered an opportunity to improve closure geometry. Rosetta chain closure involves three virtual N, C, and C$\alpha$ atoms prepended or appended to the residue immediately after or before the chainbreak, respectively; these virtual atoms should perfectly overlap with the actual N, C, and C$\alpha$ atoms in the case of exact chain closure (chain closure RMSD of zero). Here, the Rosetta *linear_chainbreak* term, equal to sums of these atom-atom deviations (in Å), was applied with a strong weight of 150.0. The inclusion of this pseudo-energy term ensured that chains closed by CCD retained or improved near-perfect geometries during full-loop minimization, with typically less than 0.01 Å deviations from perfect closure in final, lowest energy models.

In addition to the chain closure protocol described above, which enumeratively samples the first 'gap' residue and CCD-closes the rest, all models were subjected to an analogous protocol carrying out backbone grid search at the last gap residue *j*–1, and closing the chain by CCD optimization of $\phi$ and $\psi$ at the preceding residues.

Finally, closed-loop models were also prepared by combining all models for just N-terminal fragments ($i, l+1$) and all models for just C-terminal fragments ($k-1, j$); combining these fragments gave models equivalent to those from stage ($i, j$) that were subjected to the same chain closure, full-loop side-chain optimization, and full-loop torsional optimization as above. Loop closure steps based on analytical kinematic closure were also investigated but gave fewer successful closures than the CCD-plus-minimization approach above.

*(5) Clustering of models*

For a given build-up stage ($i, j$), up to 400 models were generated from each of 400 models at previous build-up stages [($i-1, j$) and ($i, j+1$)], leading to hundreds of thousands of models. Even larger numbers were generated at the final stage of full-length loop modeling due to the many routes to chain closure. However, these models typically spanned a very large range of energies, and SWA seeks to carry forward only the lowest-energy configurations at each rebuild stage. Thus all models for a given stage were collated, filtered to retain the 4000 lowest energy models, and then reclustered. The clustering followed the procedure described above, except that RMSDs were calculated over the entire rebuilt loop fragments and a clustering RMSD threshold of 0.25 Å was applied. The 400 lowest energy configurations were carried forward. In the final stage (full-length loop models), models were re-clustered with RMSD threshold 1.0 Å, and the five lowest energy models were taken as the SWA predictions.

In the SWA runs for this study's 35-loop benchmark, some settings in the loop modeling were chosen so as to match prior benchmarks. First, for proteins containing disulfide bonds, these residue-residue pairings were assumed to be known [as in prior work [22]]; those cysteine residues were modeled without protonation of the sulfur and using standard Rosetta terms that favored disulfide bonding with typical bond lengths, angles, and torsions [54]. Second, starting structures were taken directly from the previously studied benchmark set [for the 20-protein PLOP/Rosetta test set [21,22]] or prepared by addition of hydrogens with Reduce [55], removal of all side-chain atoms, and excision of loop regions. Third, the amide H atom at the C-terminal loop endpoint ($l+1$) is coupled to the loop conformation through the torsion $\phi_{l+1}$. This degree of freedom was not sampled, as 'native' backbone hydrogen atoms were present in the benchmark starting structures. However, for the five blind tests, the hydrogens were not available *a priori*; they were initially placed in the starting excised structure with Reduce [55], and since the amide H atom at position $l+1$ was not guaranteed to be in its 'native' position, $\phi_{l+1}$ was sampled during build-up of residues $l-1$ and $l$.

It was found empirically in early tests on small loops (6-9 residues) that some cases could be solved without carrying out the full SWA dynamic programming matrix, but instead by building from the N-terminal side (without C-terminal growth), by building in separate runs from the C-terminal side (without N-terminal growth), and then combining these separate solutions with chain closure to attain final models. This simplified calculation (outlined in Figure 1e) grows as O($N$) with the number of residues $N$, rather than as O($N^2$), and is analogous to the recursion used previously for RNA loop modeling [56]. For all cases in this study, this O($N$) calculation was carried out first. If the energy gap between the lowest energy model and the second lowest energy model was less than 1 $k_B T$, the calculation was assumed to not have clearly converged on a confident model, and the loop building was repeated with the full O($N^2$) calculation, except the very long 1RHD and 7CAT test cases. Overall, 18 of 40 cases were modeled with the O($N$) calculation (see Table 2).

*SWA example command lines*
The entire SWA calculation can be set up with the following Python script, available in the Rosetta subdirectory `tools/SWA_protein_python/`:

```
generate_swa_protein_dag.py -loop_start_pdb noloop_1oyc_min.pdb -native
1oyc_min.pdb -fasta 1oyc.fasta -cluster_radius 0.25 -final_number 400 -
denovo 1 -loop_res 203-213
214 -weights score12.wts -disable_sampling_of_loop_takeoff -loop_force_
Nsquared
```

here, `noloop_1oyc_min.pdb` is the starting structure. The `-native` flag is optional and specifies a reference PDB can be supplied to compute RMSD values for the loop during the calculation. `1oyc.fasta` gives the sequence of the entire protein in FASTA format:

```
>1oyc.pdb
SFVKDFKPQALGDTNLFKPIKIGNNELLHRAVIPPLTRMRALHPGNIPNRDWAVEYYTQRAQRPGTMIITEG
AFISPQAGGYDNAPGVWSEEQMVEWTKIFNAIHEKKSFVWVQLWVLGWAAFPDNLARDGLRYDSASDNVFMD
AEQEAKAKKANNPQHSLTKDEIKQYIKEYVQAAKNSIAAGADGVEIHSANGYLLNQFLDPHSNTRTDEYGGS
IENRARFTLEVVDALVEAIGHEKVGLRLSPYGVFNSMSGGAETGIVAQYAYVAGELEKRAKAGKRLAFVHLV
EPRVTNPFLTEGEGEYEGGSNDFVYSIWKGPVIRAGNFALHPEVVREEVKDKRTLIGYGRFFISNPDLVDRL
EKGLPLNKYDRDTFYQMSAHGYIDYPTYEEALKLGWDKK
```

The numbering of the loop residues takes the convention that the first amino acid in the sequence starts at 1. Note that these numbers can be offset from the starting PDB numbering (Table 1).

Simplified calculations that take O($N$) steps with the number of loop residues $N$ (see above) are setup with the same `swa_protein_dagman.py` script above, but without the last flag `–loop_force_Nsquared`. For blind tests, the ψ torsion for the starting loop residue and ϕ torsion for the ending loop residue were sampled (see above); this was accomplished by omitting the flag `-disable_sampling_of_loop_takeoff`.

Upon running `swa_protein_dagman.py`, the entire calculation workflow is laid out in one text file describing the directed acyclic graph, `protein_build.dag`. This master job file is in the standard CONDOR DAGMAN [57] format, and specifies the jobs and their dependencies, locations of CONDOR submission files for each job, and pre- and post- processing script commands. The entire workflow can then be carried out with CONDOR's `condor_dagman`. Alternatively, the workflow has also been carried out on clusters with Load Sharing Facility (bsub), Portable Batch System (qsub), and MPI job management systems. Python scripts optimized for each of these job management systems are being made available in `tools/SWA_protein_python/`, initiated with `SWA_dagman_continuous.py`.

The CONDOR submission files, each containing the appropriate Rosetta command line, are created in a `CONDOR/` subdirectory. Additional scripts specified for pre-processing CONDOR submission files (to reflect the number of models at each stage) and for post-processing all models available for a given stage by collation into single files and removal of unused files, are available in `tools/SWA_protein_python/`.

Each of the build steps described in `protein_build.dag` corresponds to a single command line using the Rosetta executable `swa_protein_main`.

(1) An example command-line for pre-packing the 1OYC loop modeling case is:

```
swa_protein_main.<exe> -database <path to
database> -rebuild -out:file:silent_struct_type binary -fasta
1oyc.fasta -n_sample 18 -nstruct 400 -cluster:radius
0.100 -extrachi_cutoff 0 -ex1 -ex2 -score:weights
score12.wts -pack_weights pack_no_hb_env_dep.wts -in:detect_disulf
false -add_peptide_plane -native 1oyc_min.pdb -superimpose_res 1-202
215-399 -fixed_res 1-202 215-399 -calc_rms_res 203-214 -jump_res 1
399 -disable_sampling_of_loop_takeoff -mute all -s1
noloop_1oyc_min.pdb -input_res1 1-202
215-399 -use_packer_instead_of_rotamer_trials -out:file:silent
REGION_215_202/START_FROM_START_PDB/region_215_202_sample.out
```

(2) An example command line that builds residue 206 onto the end of a N-terminal fragment already containing 203–205 is:

```
swa_protein_main.<exe> -database
<database> -rebuild -out:file:silent_struct_type binary -fasta
1oyc.fasta -n_sample 18 -nstruct 400 -cluster:radius
0.100 -extrachi_cutoff 0 -ex1 -ex2 -score:weights
score12.wts -pack_weights pack_no_hb_env_dep.wts -in:detect_disulf
```

```
false -add_peptide_plane -native 1oyc_min.pdb -superimpose_res 1-202
215-399 -fixed_res 1-202 215-399 -calc_rms_res 203-214 -jump_res 1
399 -disable_sampling_of_loop_takeoff -mute all -silent1
region_215_205_sample.cluster.out -tags1 S_0 -input_res1 1-205
215-399 -sample_res 205 206 -out:file:silent
REGION_215_206/START_FROM_REGION_215_205_DENOVO_S_0/region_215_206_samp
le.out
```

Here, the build is onto the lowest energy model (`S_0`) available from a previous stage that had rebuilt residues 203–205 from the N-terminal end.

(3) An example command line that builds residue 209 onto the N-terminal end of a C-terminal fragment already containing residues 210–214:

```
swa_protein_main.<exe> -database <path to
database> -rebuild -out:file:silent_struct_type binary -fasta
1oyc.fasta -n_sample 18 -nstruct 400 -cluster:radius
0.100 -extrachi_cutoff 0 -ex1 -ex2 -score:weights
score12.wts -pack_weights pack_no_hb_env_dep.wts -in:detect_disulf
false -add_peptide_plane -native 1oyc_min.pdb -superimpose_res 1-202
215-399 -fixed_res 1-202 215-399 -calc_rms_res 203-214 -jump_res 1
399 -disable_sampling_of_loop_takeoff -mute all -silent1
region_210_202_sample.cluster.out -tags1 S_2 -input_res1 1-202
210-399 -sample_res 209 210 -out:file:silent
REGION_209_202/START_FROM_REGION_210_202_DENOVO_S_2/region_209_202_samp
le.out
```

Here, the build is onto the third lowest energy model (`S_2`) available from a previous stage that had rebuilt residues 210-214 from the C-terminal end.

(4) An example command line that takes the lowest energy model from the ensemble that has built up both the N-terminal fragment 203–206 and C-terminal fragment 209–215, samples backbone degrees of freedom at residue 207, and closes the chain by cyclic coordinate descent (CCD) across residues 207 and 208:

```
swa_protein_main.<exe> -database <path to
database> -rebuild -out:file:silent_struct_type binary -fasta
1oyc.fasta -n_sample 18 -nstruct 400 -cluster:radius
0.100 -extrachi_cutoff 0 -ex1 -ex2 -score:weights
score12.wts -pack_weights pack_no_hb_env_dep.wts -in:detect_disulf
false -add_peptide_plane -native 1oyc_min.pdb -superimpose_res 1-202
215-399 -fixed_res 1-202 215-399 -calc_rms_res 203-214 -jump_res 1
399 -disable_sampling_of_loop_takeoff -mute all -silent1
region_209_206_sample.cluster.out -tags1 S_0 -input_res1 1-206
209-399 -sample_res 208 -bridge_res 207 -cutpoint_closed
207 -ccd_close -global_optimize -out:file:silent
REGION_207_206/START_FROM_REGION_209_206_CLOSE_LOOP_CCD_S_0/region_207_
```

206_sample.out

An example command line that combines the lowest energy model for N-terminal fragment 203–206 with every model for C-terminal fragment 209–215 (built separately from each the N-terminal fragment) and carries out CCD (cyclic coordinate descent) chain closure across 207 and 208:

```
swa_protein_main.<exe> —database <path to
database> -rebuild -out:file:silent_struct_type binary -fasta
1oyc.fasta -n_sample 18 -nstruct 400 -cluster:radius
0.100 -extrachi_cutoff 0 -ex1 -ex2 -score:weights
score12.wts -pack_weights pack_no_hb_env_dep.wts -in:detect_disulf
false -add_peptide_plane -native 1oyc_min.pdb -superimpose_res 1-202
215-399 -fixed_res 1-202 215-399 -calc_rms_res 203-214 -jump_res 1
399 -disable_sampling_of_loop_takeoff -mute all -silent1
region_209_202_sample.cluster.out -tags1 S_0 -input_res1 1-202
209-399 -silent2 region_215_206_sample.cluster.out -input_res2 1-206
215-399 -bridge_res 207 208 -cutpoint_closed
206 -ccd_close -global_optimize -out:file:silent
REGION_207_206/START_FROM_REGION_209_202_REGION_215_206_CLOSE_LOOP_CCD_
S_0/region_207_206_sample.out
```

(5) An example command line for clustering the lowest energy 4000 models available for the N-terminal fragment 203–205:

```
swa_protein_main.<exe> -cluster_test -silent_read_through_errors -in:fi
le:silent
REGION_215_205/start_from_region_215_204_denovo_sample.low4000.out -in:
file:silent_struct_type binary -database <path to
database> -cluster:radius 0.25 -calc_rms_res 203-214 -out:file:silent
region_215_205_sample.cluster.out -nstruct 400 -score_diff_cut
10.000 -working_res 1-205 215-399
```

*Optimization of crystallographic loops*
To help assess efficiency of conformational sampling, the all-atom Rosetta energies of crystallographic loops were obtained with two strategies. Generally, crystallographic loops contain minor steric clashes that are penalized by the Rosetta energy function, and these conformations need to be subjected to local optimization to permit comparison to *de novo* models, with the same bond lengths and angles as used in the modeling.

The first 'idealize-and-optimize' strategy mimicked that from ref. [22]. To ensure that the loop contained the same idealized bond lengths and angles as SWA or KIC modeling, the entire crystallographic structure was subjected to Rosetta-based idealization (with the *idealize* application), and the resulting idealized loop conformation was grafted into the same side-chain pre-packed structure as used in the SWA runs above. The loop and all its neighbors were subjected to combinatorial optimization by the *packer* and then all loop torsions and all side-chain torsions in the loop and surrounding residues were subjected to continuous minimization as above. As above, keeping the backbone outside the loop residue rigorously fixed requires a formal chainbreak within the loop, which remains closed during minimization due to the `linear_chainbreak` term. For each potential chainbreak location (immediately N-terminal to the loop, and after each loop residue), 20 runs were carried out. The command line used was the following:

```
swa_protein_main.<exe> –database <path to
database> -rebuild -out:file:silent_struct_type binary -fasta
1oyc.fasta -n_sample 18 -nstruct 400 -extrachi_cutoff
0 -ex1 -ex2 -score:weights score12.wts -pack_weights
pack_no_hb_env_dep.wts -in:detect_disulf
false -add_peptide_plane -native 1oyc_min.pdb -superimpose_res 1-202
215-399 -fixed_res 1-202 215-399 -calc_rms_res 203-214 -jump_res 1
399 -disable_sampling_of_loop_takeoff -silent1
region_215_202_sample.cluster.out -tags S_0 -input_res1 1-202
215-399 -cutpoint_closed 214 -global_optimize -out:file:silent
MINIMIZE_NATIVE/12/1oyc_minimize_native.out -cluster:radius 0.0 -s2
1oyc_min_idealize.pdb -input_res2 202-215 -slice_res2 202-215
```

A second 'native SWA' strategy was used to optimize the loop conformation around the crystallographic loop. In this strategy, the entire SWA calculation after the initial side-chain prepacking was repeated, but at each sampling step, models were only carried forward if their backbone RMSD to the crystallographic loop was less than 2.0 Å. In addition, Rosetta coordinate constraints at loop Cα atoms were implemented with the following Python script command line:

```
generate_CA_constraints.py 1oyc.pdb –cst_res 203-214 –coord_cst –
anchor_res 1 –fade > 1oyc_coordinate2.0.cst
```

The script is available in `tools/SWA_protein_python/`. These constraints applied a penalty for each Cα atom deviating further than 2.0 Å from the crystallographic position, rising to a maximum of 10.0 $k_BT$ for deviations of 4.0 Å; the functional form was a cubic spline with zero derivative at 2.0 Å and 4.0 Å (the *fade* function in Rosetta). These constraints were activated in SWA runs by including flags `-rmsd_screen 2.0` and `-cst_file 1oyc_coordinate2.0.cst` in `swa_protein_main` command lines above. In all tested loops, the SWA-native strategy gave models within 1.0 Å Cα RMSD to the crystallographic loop with lower energies than the idealize-and-optimize strategy (SI Figures S2 & S3); the SWA-native values are thus reported in Table S2.

*Kinematic Closure (KIC) Monte Carlo*

To permit comparison of the SWA approach to a prior state-of-the-art method, the KIC (kinematic closure) loop modeling method in Rosetta was repeated on the 20-protein PLOP/Rosetta benchmark. The following command line was used:

```
loopmodel.<exe> -database <path to database> -loops:remodel
perturb_kic -loops:refine refine_kic -loops:input_pdb
region_FINAL.out.1.pdb -in:file:native 1oyc_min.pdb -loops:loop_file
1oyc.loop -loops:max_kic_build_attempts
10000 -in:file:fullatom -out:file:fullatom -out:prefix
1oyc -out:pdb -ex1 -ex2 -ex1aro -extrachi_cutoff 0 -out:nstruct
1000 -out:file:silent_struct_type binary -out:file:silent
1oyc_kic.out -fix_ca_bond_angles -kic_use_linear_chainbreak -allow_omeg
a_move -sample_omega_at_pre_prolines
```

The command line is identical to that used previously, except for some additional terms to ensure a fair comparison to the SWA modeling above. The flag `-fix_ca_bond_angles` retains N–Cα–C bond angles at ideal values defined by Rosetta; sampling these angles did not improve accuracy in prior work [22]. The flag `-kic_use_linear_chainbreak` uses the same `linear_chainbreak` penalty for chain closure as in the SWA runs; the original chainbreak term was found to give unacceptable deviations at chainbreaks in some SWA and KIC cases. The flag `-allow_omega_move` activates minimization at ω residues and `-sample_omega_at_pre_prolines` activates the sampling of cis proline configurations during KIC Monte Carlo moves, both matching treatment of ω torsions in the SWA approach above.

      The KIC loop modeling method requires an input starting structure with a loop pre-built, and this loop defines the fixed bond lengths and angles used in the run. Rather than using the crystallographic loops [22], this study used the lowest energy model achieved in SWA modeling above. This starting point ensured that side-chains distant from the loop were in the same conformation in the KIC and SWA runs (global re-packing of those side-chains otherwise introduces noise in energy comparisons), and that exactly the same bond geometries were used in KIC and SWA runs.

# Tables

**Table 1. Loop sequences and sources for all test cases.** For statistics $N_{contact}$, $N_{out}$, $N_{SC}$, and $N_{HB}$, residues with sequence positions within two residues of each loop residue were excluded from the calculation; only non-hydrogen atoms were considered.

| PDB | Loop | N | $N_{contact}$[a] | $N_{out}$[b] | $N_{SC}$[c] | $N_{HB}$[d] | Loop sequence | Source |
|---|---|---|---|---|---|---|---|---|
| **PLOP/Rosetta benchmark** | | | | | | | | |
| 1a8d | 155-166 | 12 | 4.8 | 2.9 | 1.9 | 1.4 | DLPDKFNAYLAN | [21,22] |
| 1arb | 182-193 | 12 | 5.2 | 4.2 | 1.3 | 2.1 | WQPSGGVTEPGS | [21,22] |
| 1bhe | 121-132 | 12 | 3.4 | 2.4 | 1.1 | 1.2 | GQGGVKLQDKKV | [21,22] |
| 1bn8 | 298-309 | 12 | 3.8 | 2.5 | 0.9 | 0.8 | STSSSSYPFSYA | [21,22] |
| 1c5e | 82-93 | 12 | 2.9 | 2.4 | 1.1 | 0.9 | YEDVLWPEAASD | [21,22] |
| 1cb0 | 33-44 | 12 | 3.8 | 2.8 | 1.4 | 1.4 | YVDTPFGKPSDA | [21,22] |
| 1cnv | 188-199 | 12 | 4.3 | 3.0 | 1.3 | 1.2 | FYNDRSCQYSTG | [21,22] |
| 1cs6 | 145-156 | 12 | 2.8 | 2.7 | 1.4 | 1.1 | NEFPNFIPADGR | [21,22] |
| 1dqz | 209-220 | 12 | 3.1 | 2.6 | 0.8 | 0.8 | CGNGTPSDLGGD | [21,22] |
| 1exm | 291-302 | 12 | 4.1 | 2.6 | 1.4 | 1.3 | RGVSREEVERGQ | [21,22] |
| 1f46 | 64-75 | 12 | 2.6 | 2.2 | 1.1 | 0.6 | MVKPGTFDPEMK | [21,22] |
| 1i7p | 63-74 | 12 | 3.9 | 3.1 | 1.1 | 1.2 | LPSPQHILGLPI | [21,22] |
| 1m3s[e] | 68-79 | 12 | 2.2 | 1.8 | 0.7 | 1.2 | VGEILTPPLAEG | [21,22] |
| 1ms9 | 529-540 | 12 | 2.5 | 2.5 | 0.8 | 0.8 | GSTPVTPTGSWE | [21,22] |
| 1my7 | 254-265 | 12 | 2.9 | 2.8 | 0.8 | 0.5 | TPPYADPSLQAP | [21,22] |
| 1oth | 69-80 | 12 | 2.7 | 2.3 | 0.8 | 1.0 | QKGEYLPLLQGK | [21,22] |
| 1oyc | 203-214 | 12 | 4.9 | 3.6 | 1.9 | 1.9 | DPHSNTRTDEYG | [21,22] |
| 1qlw | 31-42 | 12 | 2.9 | 1.4 | 0.8 | 1.0 | ETLSLSPKYDAH | [21,22] |
| 1t1d | 127-138 | 12 | 3.1 | 2.4 | 0.7 | 1.1 | SGGRLRRPVNVP | [21,22] |
| 2pia | 30-41 | 12 | 2.6 | 2.1 | 0.6 | 0.9 | DPQGAPLPPFEA | [21,22] |
| **Difficult cases** | | | | | | | | |
| 1alc | 34-41 | 8 | 5.2 | 4.5 | 1.1 | 1.2 | SGYDTQAI | [21,58] |
| 1msp | 54-62 | 9 | 3.4 | 3.4 | 0.9 | 1.0 | SVDPPCGVL | [35] |
| 1w7z | 40952 | 12 | 1.9 | 1.9 | 0.8 | 0.6 | CPRILIRCKQDS | [33,59][f] |
| 2tgi | 48-59 | 12 | 3.6 | 2.9 | 1.4 | 1.8 | CPYLWSSDTQHS | [22,60] |
| 1ppn | 175-186 | 12 | 4.9 | 3.1 | 1.9 | 1.2 | NSWGTGWGENGY | [36,61][f] |
| 1bni | 75-86 | 12 | 4.2 | 2.7 | 1.4 | 1.6 | DINYTSGFRNSD | [62,63][f] |
| 2ci2 | 53-64 | 12 | 1.8 | 1.8 | 0.6 | 0.7 | VGTIVTMEYRID | [33,64,65] |
| 1udg | 152-163 | 12 | 3.8 | 2.1 | 1.2 | 1.5 | VKRGAAASHSRI | [66][f] |
| 1arp | 290-307 | 18 | 3.1 | 2.8 | 1.1 | 0.8 | LTDCSDVIPSAVSNNAAP | [34] |
| 1huw | 47-64 | 18 | 2.4 | 1.9 | 1.1 | 1.0 | NPQTSLCPSESIPTPSNK | [60] |
| 1rhd[f] | 136-153 | 18 | 2.3 | 2.3 | 0.8 | 0.7 | EGHPVTSEPSRPEPAIFK | [60] |
| 7cat | 290-309 | 20 | 3.8 | 2.4 | 1.7 | 0.9 | IFPFNPFDLTKVWPHGDYPL | [60] |
| 1thg | 309-332 | 24 | 4.8 | 3.0 | 1.3 | 1.5 | LFGLLPQFLGFGPRPDGNIIPDAA | [22,60] |
| 1c5e[f] | 70-93 | 24 | 4.2 | 3.6 | 1.1 | 1.1 | TTLTFYKSGTFRYEDVLWPEAASD | [21,22] |
| 1rhd | 136-164 | 29 | 2.8 | 2.6 | 0.8 | 0.7 | EGHPVTSEPSRPEPAIFKATLNRSLLKTY | [60] |
| **Blind tests** | | | | | | | | |
| Test A | 22-33 | 12 | 1.0 | 0.7 | 0.2 | 0.3 | NITETFPLKPGQ | Blind[g] |
| Test B | 57-68 | 12 | 2.4 | 1.8 | 0.4 | 1.2 | ISKCPIANSDPR | Blind[g] |
| Test C | 92-103 | 12 | 3.6 | 1.8 | 1.1 | 2.0 | RDFVRDSTSTNK | Blind[g] |
| Test D | 210-220 | 11 | 2.7 | 1.6 | 0.4 | 1.2 | TFVRHPEHEEA | Blind[g] |
| 3v7e | 534-545 | 12 | 3.8 | 2.2 | 0.8 | 1.8 | AKDADPILTSSV | Blind [39] |
| | Mean | 13.2 | 3.3 | 2.5 | 1.0 | 1.1 | | |
| | Median | 12.0 | 3.1 | 2.5 | 1.1 | 1.1 | | |

[a] Average number of residues that make at least one atom-atom contact (distance < 4.0 Å) with each loop residue.
[b] Avg. number of residues outside the loop that make an atom-atom contact (dist. < 4.0 Å) with each loop residue.
[c] Avg. number of residues that make an atom-atom contact (dist. < 4.0 Å) to a loop residue involving an atom requiring side-chain placement (not N, C, Cα, Cβ, O).
[d] Avg. number of hydrogen bonds per residue, defined as donor/acceptor pairs with distance less than 3.2 Å.
[e] Test included two crystallographic neighbors that interact with loop [21,22].
[f] Loop with irregular structure that remains rigid upon binding to inhibitors or protein partners; see cited references.
[g] 275-residue protein crystal structure from W. Weis and colleagues (see Figs. 5a-d).

**Table 2. Accuracy achieved on 40 loop modeling cases.**

| Target | Length | Cα RMSD to crystallographic loop (Å) | | | | | |
|---|---|---|---|---|---|---|---|
| | | Lowest RMSD model | | Best of five models (rank) | | Lowest energy model | |
| | | KIC | SWA | KIC | SWA | KIC | SWA |
| **PLOP/Rosetta benchmark** | | | | | | | |
| 1a8d[a] | 12 | 0.71 | 0.33 | 0.71 (1) | 0.42 (1) | 0.71 | 0.42 |
| 1arb[a] | 12 | 0.59 | 0.48 | 1.54 (1) | 0.48 (1) | 1.54 | 0.48 |
| 1bhe | 12 | 0.50 | 0.30 | 0.61 (1) | 0.30 (1) | 0.61 | 0.30 |
| 1bn8 | 12 | 0.51 | 0.33 | 0.92 (1) | 0.63 (2) | 0.92 | 1.27 |
| 1c5e[b] | 12 | 0.34 | 0.36 | 0.36 (1) | 0.44 (2) | 0.36 | 1.25 |
| 1cb0[a] | 12 | 0.46 | 0.32 | 0.56 (1) | 0.64 (1) | 0.56 | 0.64 |
| 1cnv | 12 | 0.84 | 1.22 | 1.59 (1) | 1.59 (1) | 1.59 | 1.59 |
| 1cs6[a] | 12 | 0.93 | 0.66 | 1.94 (4) | 0.79 (1) | 2.87 | 0.79 |
| 1dqz[a] | 12 | 0.52 | 0.38 | 0.76 (1) | 0.48 (1) | 0.76 | 0.48 |
| 1exm[a] | 12 | 0.65 | 0.48 | 0.98 (1) | 0.62 (1) | 0.98 | 0.62 |
| 1f46 | 12 | 0.50 | 0.38 | 0.57 (1) | 0.38 (1) | 0.57 | 0.38 |
| 1i7p | 12 | 0.39 | 0.39 | 0.49 (2) | 0.43 (3) | 2.83 | 1.61 |
| 1m3s[a] | 12 | 0.27 | 0.27 | 0.36 (1) | 0.27 (2) | 0.36 | 3.24 |
| 1ms9 | 12 | 0.24 | 0.34 | 0.39 (1) | 0.34 (1) | 0.39 | 0.34 |
| 1my7 | 12 | 0.35 | 0.34 | 0.75 (1) | 0.51 (1) | 0.75 | 0.51 |
| 1oth[a] | 12 | 0.31 | 0.43 | 0.39 (1) | 0.71 (1) | 0.39 | 0.71 |
| 1oyc | 12 | 1.68 | 0.38 | 3.03 (2) | 0.39 (1) | 4.53 | 0.39 |
| 1qlw | 12 | 1.00 | 0.45 | 1.24 (1) | 0.66 (3) | 1.24 | 4.98 |
| 1t1d | 12 | 0.45 | 0.31 | 0.90 (1) | 0.41 (1) | 0.90 | 0.41 |
| 2pia[a] | 12 | 0.67 | 0.55 | 1.10 (1) | 0.83 (3) | 1.10 | 1.06 |
| **Difficult cases** | | | | | | | |
| 1alc[a] | 8 | 0.17 | 0.29 | 0.25 (1) | 0.58 (2) | 0.25 | 0.7 |
| 1msp[a] | 9 | 0.27 | 0.52 | 0.55 (1) | 0.73 (1) | 0.55 | 0.73 |
| 1w7z | 12 | 0.35 | 0.40 | 0.80 (1) | 0.79 (1) | 0.80 | 0.79 |
| 2tgi | 12 | 0.70 | 0.53 | 1.57 (2) | 0.53 (3) | 2.73 | 2.87 |
| 1ppn | 12 | 0.36 | 0.44 | 0.70 (1) | 0.89 (1) | 0.70 | 0.89 |
| 1bni | 12 | 0.73 | 0.74 | 1.76 (2) | 1.10 (4) | 2.79 | 1.12 |
| 2ci2 | 12 | 0.69 | 0.82 | 2.35 (2) | 3.44 (5) | 2.73 | 4.50 |
| 1udg | 12 | 0.48 | 0.53 | 0.94 (4) | 2.43 (3) | 1.92 | 3.29 |
| 1arp[a] | 18 | 0.83 | 0.81 | 0.94 (1) | 1.60 (1) | 0.94 | 1.60 |
| 1huw[a] | 18 | 0.67 | 1.16 | 0.67 (3) | 1.35 (1) | 1.92 | 1.35 |
| 1rhd[a,b] | 18 | 0.87 | 0.74 | 0.87 (1) | 0.82 (2) | 0.87 | 1.66 |
| 7cat[a] | 20 | 4.02 | 7.18 | 6.86 (5) | 7.69 (1) | 7.65 | 7.69 |
| 1thg[a] | 24 | 3.90 | 0.74 | 5.44 (2) | 0.80 (1) | 6.88 | 0.80 |
| 1c5e[a,b] | 24 | 2.02 | 0.28 | 2.02 (2) | 0.41 (1) | 7.47 | 0.41 |
| 1rhd[a,b] | 29 | 4.67 | 3.57 | 11.28 (4) | 12.20 (5) | 16.43 | 17.63 |
| **Blind tests** | | | | | | | |
| Test A | 12 | 0.46 | 0.65 | 0.74 (1) | 0.91 (3) | 0.74 | 2.24 |
| Test B | 12 | 0.52 | 0.49 | 1.03 (1) | 0.91 (1) | 1.03 | 0.91 |
| Test C | 12 | 0.58 | 0.38 | 0.78 (2) | 0.54 (1) | 1.44 | 0.54 |
| Test D | 11 | 0.23 | 0.29 | 0.38 (1) | 0.52 (1) | 0.38 | 0.52 |
| 3v7e[a] | 12 | 0.51 | 0.42 | 0.82 (2) | 0.53 (1) | 1.62 | 0.53 |
| Mean | 13.5 | 0.87 | **0.74** | 1.50 | **1.25** | 2.09 | **1.81** |
| Median | 12 | 0.80 | **0.44** | 0.84 | **0.64** | 0.96 | **0.80** |
| RMSD < 1.0 Å | | 34/40 | **36/40** | 26/40 | **32/40** | 22/40 | **23/40** |

[a]SWA runs carried out with simplified O(*N*) calculation scheme; see methods.
[b]Longer and shorter variants of loops were modeled separately; see Table 1.

**Figure Legends**

**Figure 1. Schematics of stepwise assembly calculation**. **(a-c)** Degrees of freedom sampled by residue-level enumeration (red torsions in backbone, labeled) and by side-chain combinatorial optimization (green torsions) for addition to N-terminal fragment **(a)**, addition to C-terminal fragment **(b)**, and chain-closure step **(c)**. In (a) and (b), note presence of methylamide and acetyl 'caps', respectively, to model peptide connection to next residue. **(d)** Directed acyclic graph (DAG) outlining overall calculation. Movements leftward or downward in the graph indicate building on loop N-terminal fragment and C-terminal fragment, respectively. Each filled circle represents a stage $(i, j)$ at which models are clustered. The diagram is for a loop with $N = 6$ residues. Chain closure steps (cyan arrows) for models with one-residue gap between N- and C- terminal fragments are shown; for clarity, steps that close two- or three- residue gaps are not shown. **(e)** Simplified DAG in which fragments are built from N-terminal end without concomitant growth in C-terminal end, or vice versa, followed by chain closure. This calculation takes $O(N)$ computational expense, compared to $O(N^2)$ expense of the full DAG in (a).

**Figure 2. Stepwise assembly applied to a challenging loop prediction problem. (a-k)** Stepwise assembly (SWA) of residues 203–214 from PDB ID *1oyc*. In each panel, the added residue (carbon atoms in magenta), previously built loop residues (pink), surrounding side-chains that interact with the new residue (green), and newly formed hydrogen bonds (dashed lines) are highlighted. **(l)** Final loop (pink) from this build-up path (a-k) agrees with the crystallographic loop (blue) with atomic accuracy (Cα RMSD 0.39 Å). Surrounding side-chains are shown in white (SWA model) and pale cyan

(crystallographic model). **(m)** Energy vs. RMSD of all SWA models (red), generated by recursively following all build-up paths, compared to models from Rosetta kinematic closure Monte Carlo (gray). Arrow marks the lowest energy model discovered; this is the conformation in (k) and (l). **(n)** Dynamic-programming-style matrix highlighting the residue-by-residue build-up in path (a-k).

**Figure 3. Effect of build-up path on loop conformations**. **(a)** Five low energy conformations, and **(b)** corresponding build-up paths and Rosetta all-atom energies (numerical values given in Rosetta units, approximately 1 $k_BT$) from the *1oyc* test case of Figure 1. Different build up paths can give similar configurations (compare brown and blue loops). Similar but distinct paths can give substantially different configurations (compare green, orange, and pink loops).

**Figure 4. Sub-Angstrom accuracy in a benchmark of difficult protein loops by stepwise assembly.** Each panel overlays the best of five lowest energy models from stepwise assembly (SWA; carbon atoms in pink) on the crystallographic loop (blue). The build-up path that gave the SWA model is shown as a dynamic-programming-style matrix (black arrows mark single-residue additions; gray arrow marks chain closure step). The Cα RMSD values achieved for each puzzle (with rank of presented model among top 5 SWA models in parentheses) were: **(a)** *1f46*, 0.46 Å (1st); **(b)** *1a8d*, 0.42 Å (1st); **(c)** *1cs6*, 0.79 Å (1st); **(d)** *1qlw*, 0.66 Å (3rd); **(e)** *1msp*, 0.73 Å (1st); **(f)** *1alc*, 0.58 Å (2nd); **(g)** *1ppn*, 0.89 Å (1st); **(h)** *1c5e* (24-residue), 0.41 Å (1st); **(i)** *1thg* (24-residue), 0.80 Å (1st). In (f), the top-ranked build-up path involved generation of N-terminal and C-terminal fragments

separately (leftward and downward black arrow paths), followed by recombination and chain closure; see Methods.

**Figure 5. Sub-angstrom accuracy in blind structure prediction of protein loops.** Each panel overlays the best of five lowest energy models from stepwise assembly (SWA; carbon atoms in pink) on the crystallographic loop (blue). The build-up path that gave the SWA model is shown as a dynamic-programming-style matrix (black arrows mark single-residue additions; gray arrow marks chain closure step). **(a-d)** Recovery of loops of an unreleased structure of 275-residue protein with all loops and all side-chains removed, with Cα RMSDs of 0.91 Å, 0.91 Å, 0.54 Å, 0.52 Å. **(e)** *3v7e* RNA-puzzle (RMSD 0.54 Å), with RNA component shown in green.

**Figure 1**

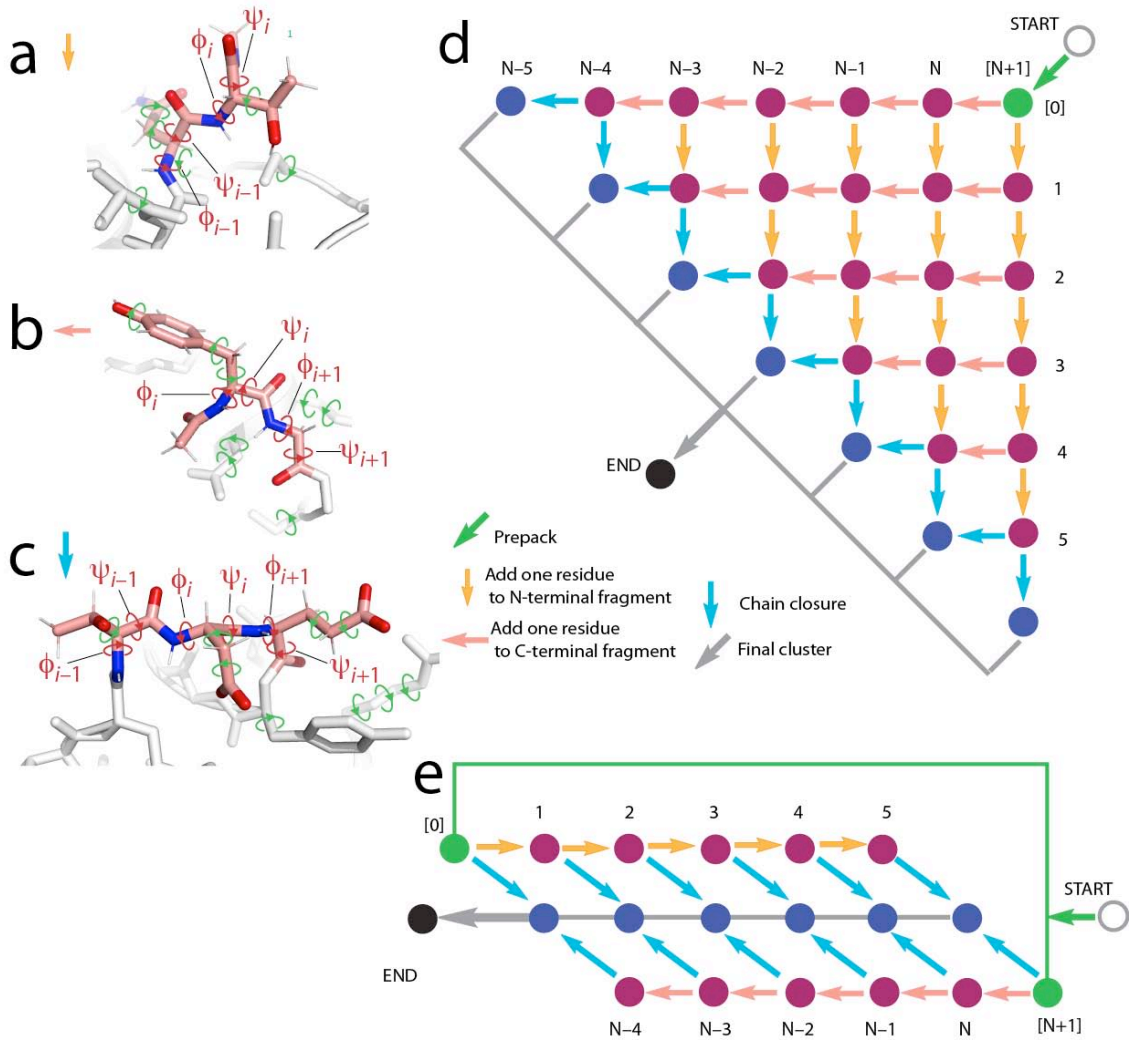

**Figure 2**

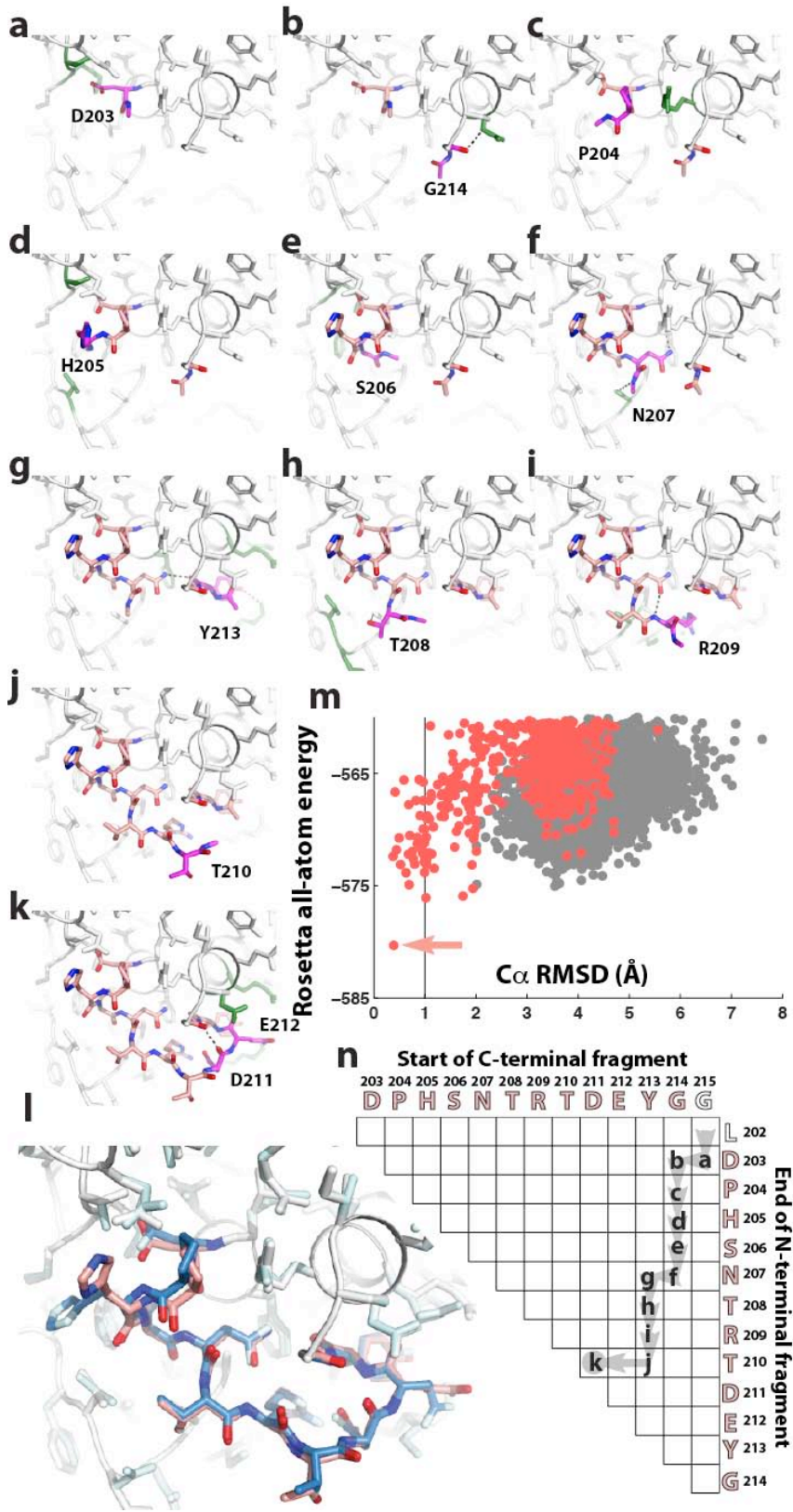

**Figure 3**

a b

−571.2
−575.3
−575.9
−576.4
−580.3

**Figure 4**

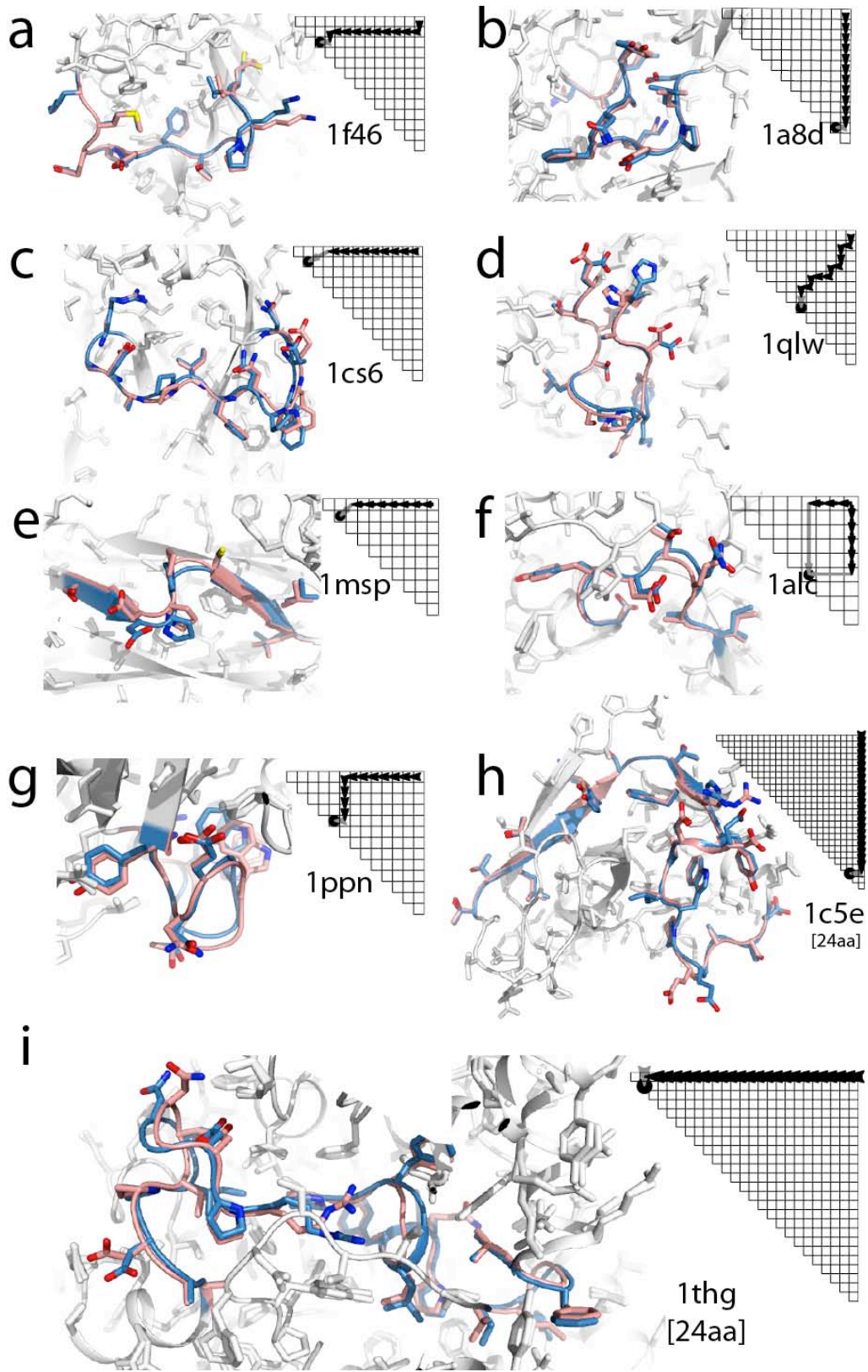

**Figure 5**

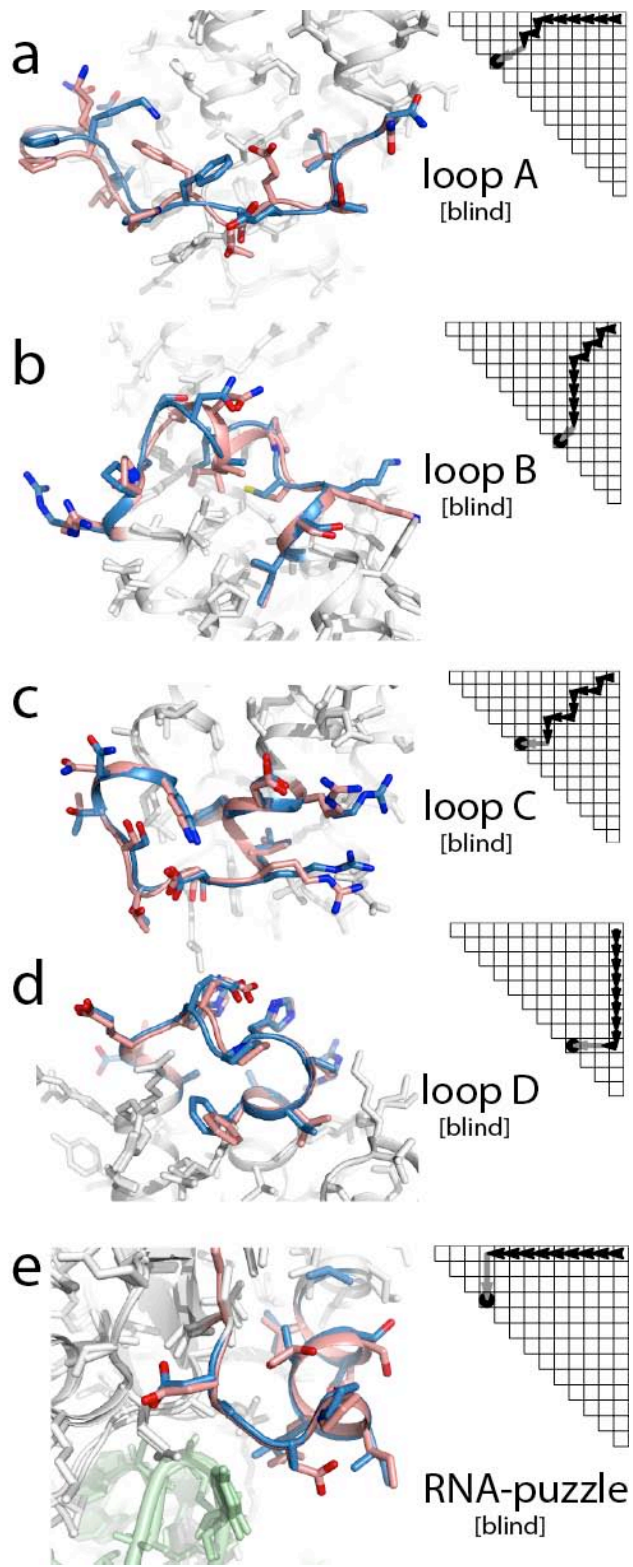

**Supporting Information for** "Atomic-accuracy prediction of protein loop structures through an RNA-inspired ansatz"

Rhiju Das, Departments of Biochemistry and Physics, Stanford University, Stanford, CA 94305, USA

This Supporting Information document contains:

**Supporting Information Table S1.** Comparison of all loop modeling methods in 20-residue PLOP/Rosetta benchmark.

**Supporting Information Table S2.** Energy comparisons to determine convergence and conformational sampling efficiency.

**Supporting Information Figure S1.** Energy vs. RMSD plots at intermediate stages of stepwise assembly build-up.

**Supporting Information Figure S2.** Energy vs. RMSD summaries of modeling runs for 20-loop PLOP/Rosetta benchmark.

**Supporting Information Figure S3.** Energy vs. RMSD summaries of modeling runs for 15 difficult and 5 blind test cases.

**Supporting References**

**Supporting Information Table S1. Comparison of all loop modeling methods in 20-residue PLOP/Rosetta benchmark.**

| Target | KIC, Number of models[a] | Cα RMSD to crystallographic loop (Å) | | | | | | |
|---|---|---|---|---|---|---|---|---|
| | | KIC, lowest RMSD | KIC, best of 5 (rank) | Comparison of lowest energy models | | | | |
| | | | | KIC | KIC[b] | Frag.Assemb.[b] | PLOP[c] | SWA |
| 1a8d | 4906 | 0.71 | 0.71 (1) | 0.71 | 6.9 | 5.4 | 2.8 | 0.42 |
| 1arb | 6635 | 0.59 | 1.54 (1) | 1.54 | 1.0 | 1.6 | 2.6 | 0.48 |
| 1bhe | 7645 | 0.50 | 0.61 (1) | 0.61 | 0.8 | 7.1 | 0.7 | 0.30 |
| 1bn8 | 6193 | 0.51 | 0.92 (1) | 0.92 | 0.7 | 2.5 | 2.6 | 1.27 |
| 1c5e | 12198 | 0.34 | 0.36 (1) | 0.36 | 0.5 | 0.8 | 1.7 | 1.25 |
| 1cb0 | 9302 | 0.46 | 0.56 (1) | 0.56 | 0.6 | 1.0 | 0.3 | 0.64 |
| 1cnv | 6879 | 0.84 | 1.59 (1) | 1.59 | 1.4 | 2.3 | 3.3 | 1.59 |
| 1cs6 | 7513 | 0.93 | 1.94 (4) | 2.87 | 3.0 | 2.5 | 3.5 | 0.79 |
| 1dqz | 8525 | 0.52 | 0.76 (1) | 0.76 | 0.7 | 1.9 | 0.6 | 0.48 |
| 1exm | 6019 | 0.65 | 0.98 (1) | 0.98 | 0.9 | 0.6 | 0.5 | 0.62 |
| 1f46 | 14228 | 0.50 | 0.57 (1) | 0.57 | 2.5 | 2.1 | 1.1 | 0.38 |
| 1i7p | 7119 | 0.39 | 0.49 (2) | 2.83 | 2.7 | 0.7 | 0.3 | 1.61 |
| 1m3s[d] | 11730 | 0.27 | 0.36 (1) | 0.36 | 6.3 | 3.6 | 5.6 | 3.24 |
| 1ms9 | 4712 | 0.24 | 0.39 (1) | 0.39 | 0.4 | 2.5 | 2.5 | 0.34 |
| 1my7 | 9451 | 0.35 | 0.75 (1) | 0.75 | 2.3 | 2.0 | 0.9 | 0.51 |
| 1oth | 8567 | 0.31 | 0.39 (1) | 0.39 | 0.6 | 0.6 | 0.7 | 0.71 |
| 1oyc | 16878 | 1.68 | 3.03 (2) | 4.53 | 4.0 | 3.2 | 1.2 | 0.39 |
| 1qlw | 5860 | 1.00 | 1.24 (1) | 1.24 | 1.0 | 3.3 | 1.4 | 4.98 |
| 1t1d | 10378 | 0.45 | 0.90 (1) | 0.90 | 0.8 | 0.5 | 1.0 | 0.41 |
| 2pia | 6479 | 0.67 | 1.10 (1) | 1.10 | 1.0 | 1.1 | 0.5 | 1.06 |
| Mean | 8053 | 0.60 | 0.91 (1) | 1.15 | 1.9 | 2.3 | 1.7 | **1.07** |
| Median | 7316 | 0.51 | 0.76 (1) | 0.83 | 1.0 | 2.1 | 1.2 | **0.63** |
| RMSD < 1.0 Å | | 18 | 14 | 13 | 10 | 6 | 8 | **14** |

[a] Results from applying 6400 CPU-hours of computation (Intel Xeon E5345 2.33 GHz). For 1OYC case, 16000 CPU-hours were expended.
[b] From Supplementary Table 2 in ref. (1). Number of KIC models in previous study was 1000.
[c] From Supplementary Table S4 ("new protocol") in ref. (2).
[d] SWA and repeated KIC runs included two crystallographic neighbors that interact with loop (1, 2).

**Supporting Information Table S2. Energy comparisons to estimate convergence and conformational sampling efficiency.**

| Target | Cα RMSD Best of 5 (rank) | All–atom Rosetta Energy differences[c] | | |
|---|---|---|---|---|
| | | SWA: (2) – (1)[d] | SWA –NATIVE | SWA – KIC |
| 1a8d[a] | 0.42 (1) | 8.8 | –0.4 | –4.3 |
| 1arb[a] | 0.48 (1) | 5.5 | –0.2 | –1.6 |
| 1bhe | 0.30 (1) | 2.9 | 2.2 | –5.5 |
| 1bn8 | 0.63 (2) | 0.8 | –1.2 | –2.0 |
| 1c5e | 0.44 (2) | 0.3 | 0.4 | 1.1 |
| 1cb0[a] | 0.64 (1) | 8.4 | 0.5 | –0.6 |
| 1cnv | 1.59 (1) | 5.7 | 0.0 | –5.5 |
| 1cs6[a] | 0.79 (1) | 0.7 | 0.8 | –2.1 |
| 1dqz[a] | 0.48 (1) | 3.7 | 1.8 | –0.0 |
| 1exm[a] | 0.62 (1) | 4.9 | 2.4 | 0.3 |
| 1f46 | 0.38 (1) | 4.0 | –3.0 | –1.1 |
| 1i7p | 0.43 (3) | 1.5 | 0.0 | –4.4 |
| 1m3s[a,g] | 0.27 (2) | 1.6 | 0.5 | 4.5 |
| 1ms9 | 0.34 (1) | 0.5 | –0.4 | –0.2 |
| 1my7 | 0.51 (1) | 0.7 | 0.3 | 1.5 |
| 1oth[a] | 0.71 (1) | 3.5 | 1.4 | –1.5 |
| 1oyc | 0.39 (1) | 4.4 | 1.4 | –2.6 |
| 1qlw | 0.66 (3) | 0.4 | 3.0 | –0.1 |
| 1t1d | 0.41 (1) | 0.6 | 0.5 | 0.4 |
| 2pia[a] | 0.83 (3) | 7.0 | –0.1 | –1.8 |
| 1alc[a] | 0.58 (2) | 3.1 | 0.3 | -8.1 |
| 1msp[a] | 0.73 (1) | 7.0 | 0.3 | 11.6 |
| 1w7z | 0.79 (1) | 1.8 | –0.1 | -9 |
| 2tgi | 0.53 (3) | 1.6 | –3.4 | -12 |
| 1ppn | 0.89 (1) | 1.0 | –1.3 | -9.1 |
| 1bni | 1.10 (4) | 1.4 | 2.9 | 1.4 |
| 2ci2 | 3.44 (5) | 0.0 | –0.9 | 4 |
| 1udg | 2.43 (3) | 1.7 | 0.8 | 3.3 |
| 1arp[a] | 1.60 (1) | 0.5 | –0.7 | -11.5 |
| 1huw[a] | 1.35 (1) | 4.2 | 1.7 | -11.8 |
| 1rhd[a] | 0.82 (2) | 0.8 | 4.5 | 4.1 |
| 7cat[a] | 7.69 (1) | 0.1 | 7.5 | 6.1 |
| 1thg[a] | 0.80 (1) | 4.1 | 12.3 | –26.2 |
| 1c5e[a,b] | 0.41 (1) | 1.8 | 1.4 | –14.8 |
| 1rhd[a,b] | 12.20 (5) | 0.5 | 14.2 | -0.5 |
| testA | 0.91 (3) | 0.4 | –1.7 | 0.3 |
| testB | 0.91 (1) | 0.8 | –5.7 | -6.1 |
| testC | 0.54 (1) | 1.3 | –5.1 | -0.9 |
| testD | 0.52 (1) | 2.7 | –1.5 | -0.8 |
| 3v7e[a] | 0.53 (1) | 3.7 | n.d.[e] | n.d.[e] |
| Mean | 1.27 (1) | 2.4 | 0.9 | –2.7[f] |
| Median | 0.64 (1) | 1.6 | 0.3 | –1.1[f] |

[a] SWA runs carried out with simplified O(N) calculation scheme; see Supporting Information methods.
[b] Longer and shorter variants of loops were modeled separately; see Supporting Information Table S1.
[c] Energies given in Rosetta units (approximately 1 $k_BT$).
[d] Energy difference between second lowest energy SWA model and lowest energy SWA model, used to assess convergence.
[e] For 3v7e (blind test, RNA-binding protein ybxF), optimized crystallographic energies cannot be compared to loops built on comparative model due to differences in starting scaffold.
[g] Test included two crystallographic neighbors that interact with loop (1, 2).

**Supporting Information Figure S1. Energy vs. RMSD plots at intermediate stages of stepwise assembly build-up**. Build-up in panels (a) to (k) corresponds to residue-by-residue path in Figs. 1a-k of main text for the *1oyc* loop (residues 203–214). RMSD over all backbone heavy-atoms (N, C, Cα, and O) is shown on x-axis, using the corresponding loop fragments in the crystallographic loop as a reference. In each panel, symbol with black outline marks the specific model that eventually leads to the final lowest energy model in (k).

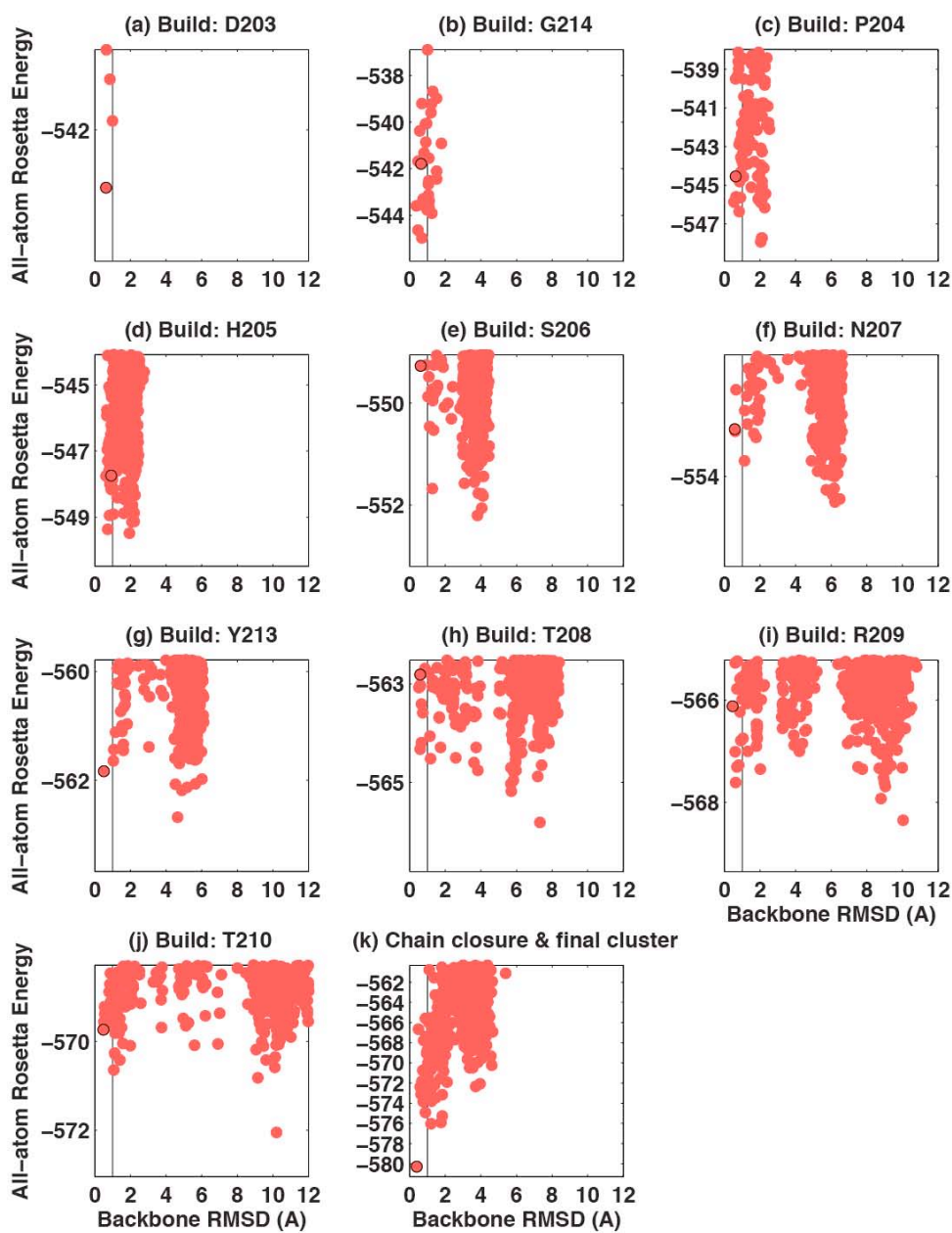

**Supporting Information Figure S2. Energy vs. RMSD summaries of modeling runs for 20-loop PLOP/Rosetta benchmark**. Rosetta all-atom energy values and loop Cα RMSDs are plotted for models from kinematic closure Monte Carlo (KIC, gray); stepwise assembly with O($N$) simple calculation (red); stepwise assembly with full O($N^2$) build-up (pink); crystallographic loops optimized by SWA re-building with constraints (blue); and crystallographic loops optimized by idealizing, re-packing, and continuous minimization (green).

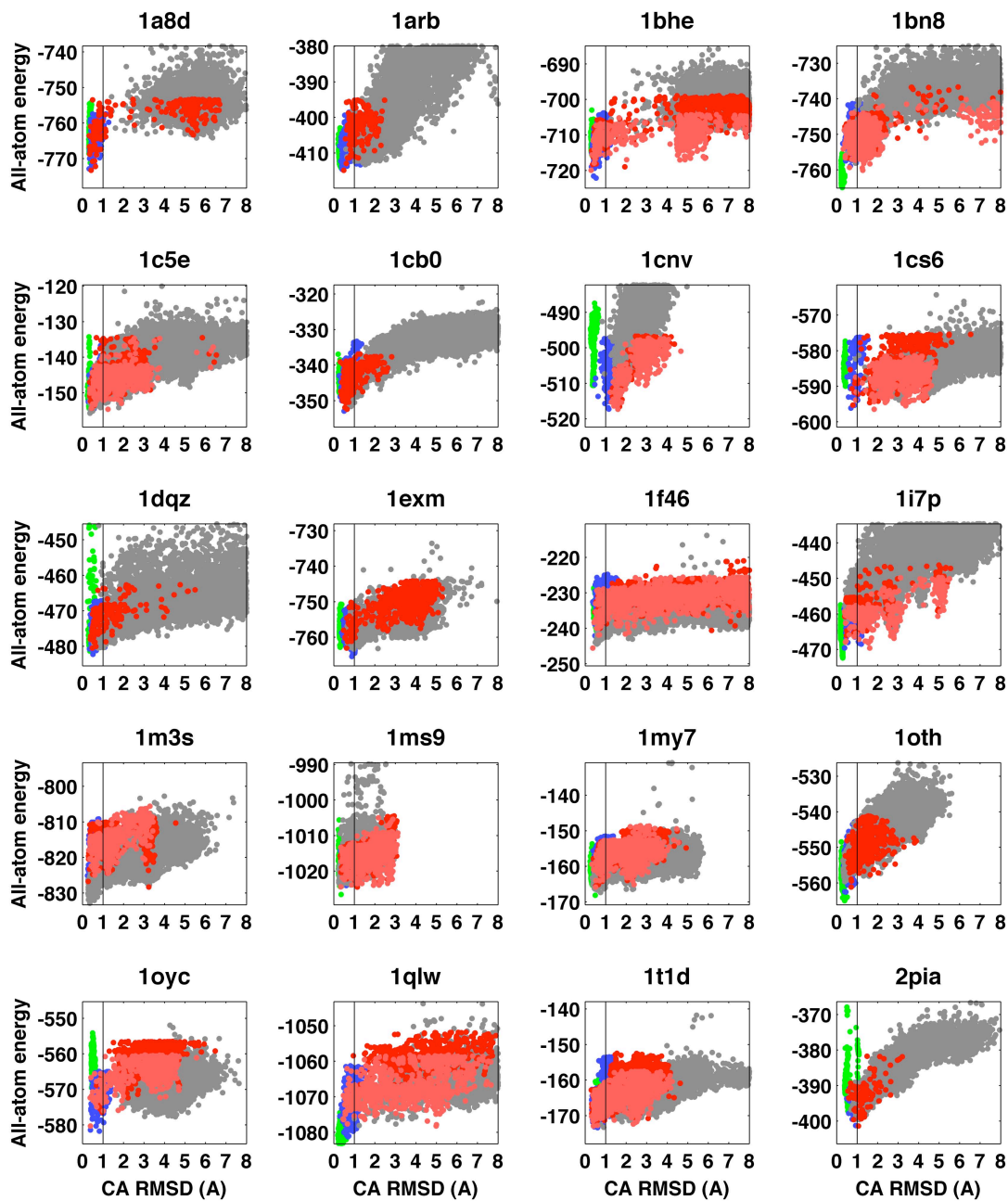

**Supporting Information Figure S3. Energy vs. RMSD summaries of modeling runs for 15 difficult and 5 blind test cases**. Rosetta all-atom energy values and loop Cα RMSDs are plotted for models from models from kinematic closure Monte Carlo (KIC, gray); stepwise assembly with O(*N*) simple calculation (red); stepwise assembly with full O($N^2$) build-up (pink); crystallographic loops optimized by SWA rebuilding with constraints (blue); and crystallographic loops optimized by idealizing, re-packing, and continuous minimization (green). For 3v7e (blind test, RNA-binding protein ybxF), optimized crystallographic energies are not presented, since loop building was carried out on a comparative model; and energies between KIC and SWA cannot be compared as RNA was not included in the former case.

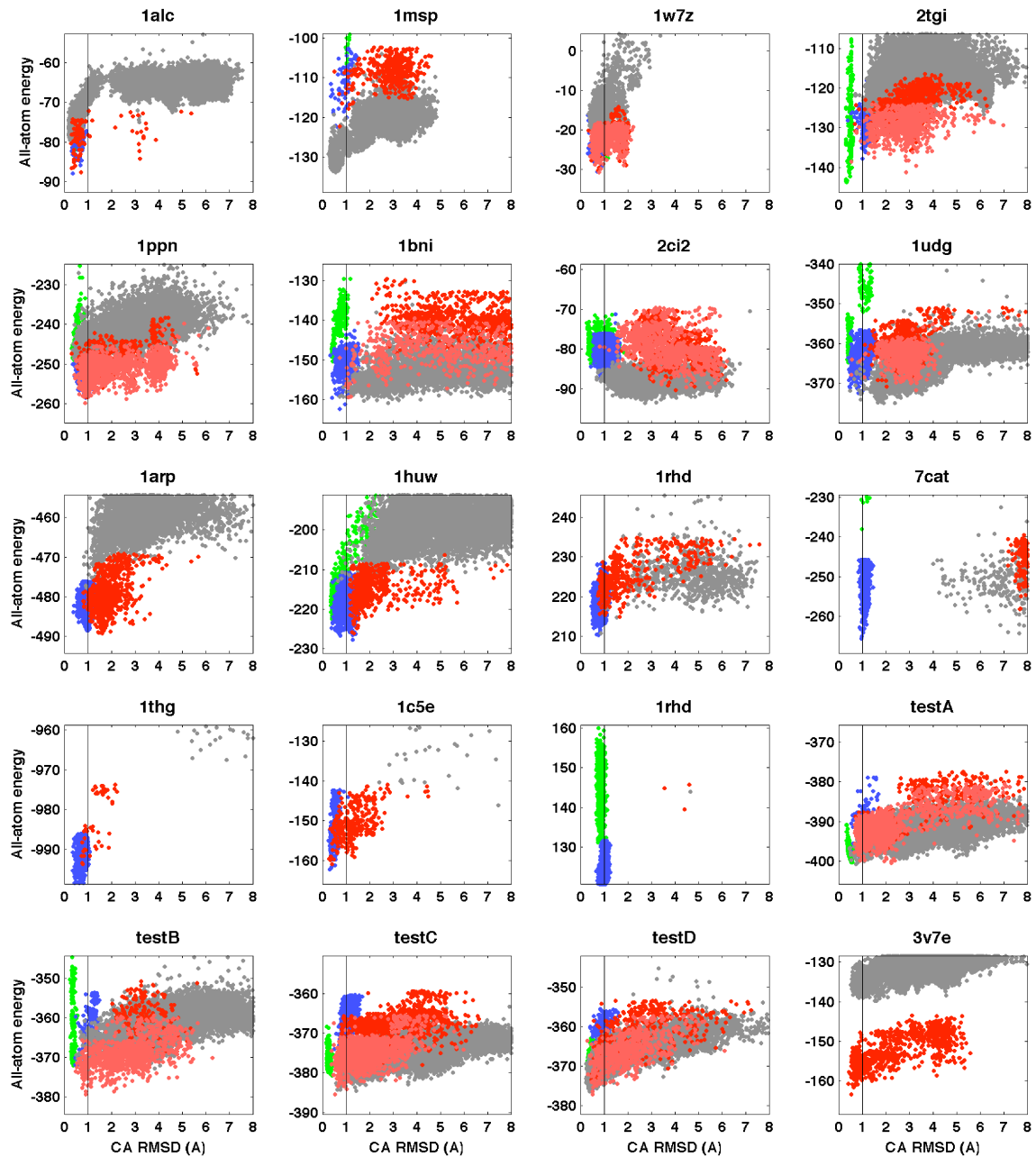

## Supporting References